\documentclass[12pt]{elsarticle}
\journal{Journal of Computational Physics}

\usepackage{mytex}
\usepackage{eepic}

\begin{document}


\title{Identification of an Optimal Derivatives Approximation by Variational Data Assimilation. }

\author{ Eugene Kazantsev}

\address{
 INRIA, projet MOISE, 
 Laboratoire Jean Kuntzmann,\\
BP 53,
38041 Grenoble Cedex 9, 
France \\
\begin{tabular}{ll}
Telephone:&+33 4 76 51 42 65 \\
Fax:& +33 4 76 63 12 63\\
\end{tabular} }
\ead{ kazan@imag.fr}

\begin{abstract} 
Variational data assimilation technique applied to  identification of   optimal approximations of derivatives near  boundary  is discussed  in frames of one-dimensional wave equation. Simplicity of the equation and of its numerical scheme allows us to discuss in detail as the development of  the adjoint model and assimilation results. It is shown what kind of errors can be corrected by this control and how these errors are corrected. This study is carried out   in view of using this  control to identify optimal numerical schemes in coastal regions of ocean models. 
\end{abstract}

\begin{keyword} 
Variational Data Assimilation, Boundary conditions; Wave equation

PACS: 47.85.L
\end{keyword}
\maketitle
\section{Introduction.}

It is now well known, even the best model is not sufficient to make a good forecast. Any model depends on a number of parameters, requires initial and boundary conditions and other data that must be collected and used in the model.  However, interpolating or smoothing  observed data is not the best way to incorporate these data in a model. Lorenz, in his  pioneer work  \cite{Lor63} has shown that a geophysical fluid is extremely sensitive to initial conditions. This fact requires to bring  the model and its initial data together, in order to work with the couple "model-data" and to identify the optimal initial data for the model taking into account simultaneously the information contained in the observational data and in the equations of the model.

Optimal control methods \cite{Lions68} and perturbations theory   \cite{Marchuk75} applied to the data assimilation technique (\cite{Ledimet82}, \cite{ldt86}) show the way to do it. They allow to retrieve an optimal initial point for a given model from heterogeneous observation fields. Since the early 1990's, many mathematical and geophysical teams have been involved in the development of the data assimilation strategy. One can cite many papers devoted to this problem, as  in the domain of development of different techniques for the  data assimilation  and in the domain of its applications to the atmosphere and oceans.  

However,  overwhelming majority of data assimilation methods are now intended to identify and reconstruct an optimal initial point for the model. Since Lorenz \cite{Lor63}, who has pointed out  the importance of precise knowledge of the starting point of the model, essentially  the  starting point is considered as the control parameter and the target of the data assimilation. 
 
Of course, the model's flow  is extremely sensitive to its initial point. But, it is reasonable to suppose that a geophysical  model is also sensitive to many other parameters like bottom topography, boundary conditions on rigid and open boundaries, forcing fields and friction coefficients. All  these parameters and values  are also extracted in some way  from observational data, interpolated to the model's grid and can neither be considered as exact, nor as optimal to the model. On the other hand, due to non-linearity and intrinsic instability of model's trajectory, its sensitivity to all these external parameters may also be exponential. 

Numerous studies show strong dependence of the model's flow on the boundary data (\cite{VerronBlayo}, \cite{Adcroft98}), on the representation of the bottom topography (\cite{Holland73}, \cite{EbyHolloway}, \cite{LoschHeimbach}), on the wind stress (\cite{Bryan95}, \cite{Milliff98}), on diffusivity coefficients (\cite{Bryan87}) and on fundamental parametrization like Boussinesq and hydrostatic hypotheses \cite{LoschAdcroft}.  But few papers are devoted to the development of data assimilation techniques intended  to identify and to control  these model's parameters. One can cite several attempts to use data assimilation in order to identify the bottom topography of simple models (\cite{LoschWunsch}, \cite{assimtopo}) and in order to control open boundary conditions in coastal and regional models (\cite{shulman97}, \cite{shulman98}, \cite{Taillandier}). Boundary conditions on rigid boundaries have been controlled by data assimilaton for heat equation (see for example \cite{ChenLin}, \cite{GillijnsDeMoor}), but this control concerns the diffusion operator rather than transport and advection type operators used in geophysical models.

This paper presents a preliminary study of  using  variational   data assimilation  in order to identify an optimal parametrization of boundary flows and boundary conditions on rigid boundaries.  
Despite the boundary configuration of the ocean is steady and can be measured with much better accuracy than the model's initial state, it is not obvious how to represent it  on the model's grid because of  limited resolution. The coastal line of continents possesses a very fine structure and can only be roughly approximated by the model's grid. Consequently,  boundary conditions  are defined at  the model grid's points which are different from the coast. Even the most evident impermeability condition being placed at a wrong point may lead to some error in the model's solution. From physical point of view, we should accept the flux can cross the boundary in places where the boundary is in water, prescribing some integral properties on the flux.  

Even in a fine resolution model when  boundary currents are explicitly resolved, it is not clear what kind of boundary conditions  to prescribe for tangential velocities. However, prescribing  slip or no-slip conditions may result in a drastic change of  the global circulation (see \cite{VerronBlayo}).

Consequently, it may be reasonable to use variational data assimilation in order to determine what boundary conditions are optimal for the model's variables.  However, instead of controlling boundary conditions themself, it may be more useful to  identify optimal discretization of differential operators in points adjacent to  boundaries because this is  more general case. Indeed, boundary conditions participate  in discretized operators, but considering the discretization itself, we take into account additional parameters like the position of the boundary,  lack of resolution of the grid, etc.  

In this paper we use data assimilation to control the discretization of derivatives in adjacent to boundary grid points. The development of the data assimilation is illustrated on the example of the simplest one-dimensional  wave equation. On one hand, the simplicity of the equation allows us to clearly see technical points of the development (like the algorithm of differentiation and  development of the adjoint equation) without being overwhelmed by complexity of  operators and grids. On the other hand, the knowledge of the  exact solution and of the  errors of  numerical discretization of the wave equation allow us to clearly see how these errors are corrected by data assimilation. The purpose of the paper is to study the possibility to control boundary numerical scheme by data assimilation  and the particularities of this type of control in view to develop and use the data assimilation to identify optimal numerical scheme in coastal regions of ocean models.

The paper is organized as follows. The second section describes the model, its adjoint and the data assimilation procedure. The third section is devoted to numerical experiments and discussion.

\section{One-dimensional wave equation}

As it has been noted in the introduction, we consider one-dimensional wave equation written for $u=u(x,t)$ and $p=p(x,t)$ in the following way:
\beqr
\der{u}{t} &-& \der{p}{x}=0 \nonumber \\
\der{p}{t} &-& \der{u}{x}=0 \label{wave} 
\eeqr
This equation is defined on the interval $0<x<1$ with boundary conditions prescribed for $u$ only:
\beq u(0,t)=u(1,t)=0 \label{bc}\eeq
Initial conditions are prescribed for both $u$ and $p$
\beq u(x,0)=\bar u, \; p(x,0) = \bar p \label{ic} \eeq

The equation is discretized on a regular grid that is somewhat similar to Arakawa's C grid in two dimensions:
\beqr
u_i&=&u(i h) \mbox{ for } i=0,\ldots N \nonumber \\
p_{i-1/2}&=&p(ih-h/2) \mbox{ for } i=1,\ldots N 
\eeqr
with $h=\fr{1}{N}$. This grid is well adapted to the prescribed boundary conditions because the  boundary points $x=0$ and $x=1$ belong to the grid for $u$ discretization, but do not belong to $p$-grid. 

\begin{figure}[h]
\setlength{\unitlength}{1.0mm}
\newcount\indi
\newcount\num

\begin{picture}(150,20)
\Thicklines
\put(5,10){\line(1,0){140}}
\put(5,8){\line(0,1){4}}
\put(145,8){\line(0,1){4}}
\thicklines

\multiput(5,9)(15,0){4}{\line(0,1){2}}
\indi=-1
\multiput(2,6)(15,0){4}{ 
\global\advance\indi by 1 $u_{\the\indi}$ }

\multiput(130,9)(-15,0){3}{\line(0,1){2}}
\put(144,6){ $u_{N}$ }
\indi=0
\multiput(127,6)(-15,0){3}{ 
\global\advance\indi by 1 $u_{N-\the\indi}$ }

\multiput(12,9)(15,0){3}{\line(1,1){2}}
\multiput(14,9)(15,0){3}{\line(-1,1){2}}
\indi=-1
\multiput(10,13)(15,0){3}{ 
\global\advance\indi by 1 \num=\indi \multiply\num by 2 \global\advance\num by 1 $p_{\the\num/2}$ }

\multiput(137,9)(-15,0){3}{\line(1,1){2}}
\multiput(139,9)(-15,0){3}{\line(-1,1){2}}
\indi=-1
\multiput(132,13)(-15,0){3}{ 
\global\advance\indi by 1 \num=\indi \multiply\num by 2 \global\advance\num by 1 $p_{N-\the\num/2}$ }
\end{picture} 

\end{figure}

Discrete derivatives  of $u$ and  $p$ are defined as follows
\beqr
\biggl(\der{p}{x}\biggr)_i&=&\fr{1}{h}\sum\limits_{j=-1}^{2} a_j p_{i+j-1/2} \nonumber \\
\biggl(\der{u}{x}\biggr)_{i+1/2}&=&\fr{1}{h}\sum\limits_{j=-1}^{2} a_j u_{i+j} \label{intrnlsch} 
\eeqr
at all internal points in the interval, i.e. $2\leq i \leq N-2$ for $\biggl(\der{p}{x}\biggr)_i$ and $1\leq i \leq N-2$ for $\biggl(\der{u}{x}\biggr)_{i+1/2}$. Coefficients $a_j$ are supposed to be known because we  intend to control  approximations near the boundary only. In this paper, we use either the sequence $ a_j=(0,-1,1,0)$ or  the sequence $ a_j=\fr{1}{24}(1,-27,27,-1)$ for $ j=(-1,0,1,2)$. One can easily see that corresponding approximations are of   second and of  fourth order approximation
\beqr 
\fr{p_{i+1/2}-p_{i-1/2}}{h}&=&\biggl(\der{p}{x}\biggr)_i + \fr{h^2}{24} \biggl(\tder{p}{x}\biggr)_i+ O(h^3) \nonumber\\
\fr{p_{i-3/2}-27p_{i-1/2}+27p_{i+1/2}-p_{i+3/2}}{24h}&=&\biggl(\der{p}{x}\biggr)_i - \fr{3 h^4}{640} \biggl(\pder{p}{x}\biggr)_i 
+ O(h^5)\eeqr

To be able to solve numerically the equation \rf{wave}, we need also to approximate derivatives of $u$ and $p$ near boundaries at points $i=1/2,N-1/2$ and $i=1,N-1$ respectively. These approximations are supposed to be different from \rf{intrnlsch} and include the control variables in this study. Moreover, expressions \rf{intrnlsch} can not be used at all for the fourth order approximation because they require function's values  beyond the boundary: $u_{-1}$ and $p_{-1/2}$. We can, of course, extrapolate $u$ and $p$ beyond the domain with the necessary order and substitute  extrapolated values in \rf{intrnlsch}, but it is not obvious what extrapolation formula is the best for this purpose, especially for $p$. So, in order to obtain an optimal boundary approximation assimilating external data, we suppose nothing about derivatives near the boundary points and  write them in a general form
\beqr
\biggl(\der{p}{x}\biggr)_1&=&\fr{1}{h}\sum\limits_{j=0}^{J} \alpha^p_j p_{j+1/2} \nonumber \\
\biggl(\der{u}{x}\biggr)_{1/2}&=&\fr{1}{h}\sum\limits_{j=0}^{J} \alpha^u_j u_{j} \label{bndsch} 
\eeqr
We do not fix the value of $J$ in these formula intentionally because we shall see further its influence. 

Here we can emphasize the choice to control the numerical scheme in the boundary region rather than boundary conditions.  The general form of boundary conditions that may be prescribed for $u$ variable of the one dimensional wave equation writes
$$ u(0,t)-A\der{u}{x}(0,t)=B.$$ We can not impose more complex boundary conditions (with second derivatives, for example) because we obtain a system with no solution at all.
Consequently, we can control only two parameters, $A$ and $B$. It may be sufficient in particular cases, but, as we shall see further, is not sufficient in general. However, controlling all coefficients  of  the numerical scheme \rf{bndsch}, we are free to choose as many $\alpha_j$ as we need defining appropriate value of  the parameter $J$.

We distinguish $\alpha^p$ and  $\alpha^u$ allowing different derivatives approximations for $p$ and for $u$  because of the different nature of these two functions and different boundary conditions prescribed for them.  Derivatives at the opposite side are calculated by  
\beqr
\biggl(\der{p}{x}\biggr)_{N-1}&=&-\fr{1}{h}\sum\limits_{j=0}^{J} \tilde\alpha^p_j p_{N-j-1/2} \nonumber \\
\biggl(\der{u}{x}\biggr)_{N-1/2}&=&-\fr{1}{h}\sum\limits_{j=0}^{J} \tilde\alpha^u_j u_{N-j} \label{bndschopp} 
\eeqr
and coefficients $\tilde\alpha^p$ and $\tilde\alpha^u$ are also considered as unknown control parameters different from  $\alpha^p$ and  $\alpha^u$. All together, we have $4(J+1)$ control parameters.

Time stepping is performed by leap-frog scheme
\beq
\fr{u_i^{n+1}-u_i^{n-1}}{2\tau} - \biggl(\der{p}{x}\biggr)^n_i =0, \;\; 
\fr{p_{i-1/2}^{n+1}-p_{i-1/2}^{n-1}}{2\tau} -\biggl(\der{u}{x}\biggr)^n_{i-1/2}=0 \label{timestep}
\eeq
The first time step is splitted into two stages in order to ensure  second order approximation in time and to avoid typical leap-frog  splitting between odd and even timesteps. 
\beqr 
&&\fr{u_i^{1/2}-u_i^{0}}{\tau/2} - \biggl(\der{p}{x}\biggr)^0_i =0, \;\fr{u_i^{1}-u_i^{0}}{\tau} - \biggl(\der{p}{x}\biggr)^{1/2}_i =0,
\nonumber \\
&&\fr{p_{i-1/2}^{1/2}-p_{i-1/2}^{0}}{\tau/2} - \biggl(\der{u}{x}\biggr)^0_{i-1/2} =0, \;\fr{p_{i-1/2}^{1}-p_{i-1/2}^{0}}{\tau} - \biggl(\der{u}{x}\biggr)^{1/2}_{i-1/2} =0.
 \label{fststep}
\eeqr

Approximation of the derivative introduced by \rf{intrnlsch} and \rf{bndsch} depends on control variables $\alpha$. The operator is not completely defined as in usual schemes, but it is allowed to change its properties near boundaries in order to find the best fit with requirements of the model and data.   To assign  variables  $\alpha^p$ and  $\alpha^u$ we shall perform data assimilation procedure and find their optimal values. 

\subsection{Tangent and adjoint equations}

First of all, we calculate  the Gateaux derivative of the operator with respect to control parameters. Control variables are supposed  to have small variations and we determine how these variations will perturb the solution of the model. Thus, we suppose that all $\alpha$ are replaced by some $\alpha +\delta\alpha$ such that  $\norme{\delta\alpha} << \norme{\alpha}$. Let the model with $\alpha +\delta\alpha$  have a new solution $u+\delta u, \; p+\delta p$. In this case,  variables  $\delta u, \; \delta p$ must satisfy 
 \beqr
 \der{\delta u}{t} &-& D^{(p)}\delta p -\delta D^{(p)} p -\delta D^{(p)}\delta p =0 \nonumber \\
\der{\delta p}{t} &-&D^{(u)}\delta u -\delta D^{(u)} u -\delta D^{(u)}\delta u  =0 \label{prt} 
\eeqr
where operators $D^{(p)}(\alpha^p)$ and $D^{(u)}(\alpha^u)$ are approximations of derivatives defined by \rf{intrnlsch},  \rf{bndsch}, and \rf{bndschopp}, i.e. for the $p$ derivative
\beq
D^{(p)}(\alpha^p) = \fr{1}{h}\left(
\begin{array}{ccccccccc}
\alpha^p_0&\alpha^p_1&\alpha^p_2&\alpha^p_3 & \cdots & 0& 0&0&0 \\
a_{-1}    &a_{0}     &a_{1}     &a_{2}      & \cdots & 0& 0&0&0 \\
0         &a_{-1}    &a_{0}     &a_{1}      & \cdots & 0& 0&0&0 \\
\cdots    &          &          &           & \cdots &  &  & &\cdots  \\
0         &0         &0         &0          & \cdots & a_{0} &a_{1} &a_{2}& 0  \\
0         &0         &0         &0          & \cdots & a_{-1}  &  a_{0} &a_{1} &a_{2}  \\
0         &0         &0         &0          & \cdots & -\tilde\alpha^p_3 &  -\tilde\alpha^p_2 &-\tilde\alpha^p_1 &-\tilde\alpha^p_0
\end{array}\right) \label{D}
\eeq 
Operators  $\delta D^{(p)}$ and $\delta D^{(u)}$ are the differences
\beq
\delta D^{(p)} =  D^{(p)}(\alpha^p+\delta\alpha^p)- D^{(p)}(\alpha^p) = \fr{1}{h}
\left(\begin{array}{ccccc}
\delta\alpha^p_0&\delta\alpha^p_1& \cdots & 0&0 \\
0               &0               & \cdots & 0&0 \\
\cdots          &                & \cdots &  &  \cdots  \\
0               &0               & \cdots & 0&0 \\
0               &0               & \cdots &-\delta\tilde\alpha^p_1 &-\delta\tilde\alpha^p_0
\end{array}\right)    
\eeq
and similarly for operators $\delta D^{(u)}$ and $ D^{(u)}$. 

However,  expressions $\delta D^{(p)} p$ and $\delta D^{(u)} u$ in \rf{prt} are not convenient to make further development. Writing an adjoint operator, we would better have a constant operator, which does not depend on $\delta\alpha$,   multiplied by a variable  vector which depends on $\delta\alpha$. This is the case in  products $D^{(p)}\delta p$ where $\delta p$ depends on $\delta\alpha$, but it is not the case in products like  $\delta D^{(p)} p$ where $p$ is solution of original equation and has no relation with $\delta\alpha$. It would be more convenient to  rewrite these products:
\beq
 \delta D^{(p)} p = 
 \fr{1}{h}\left(\begin{array}{c} 
 \sum\limits_{j=0}^{J} \delta\alpha^p_j p_{j+1/2} \\
 0 \\
 \vdots \\
 0 \\
-\sum\limits_{j=0}^{J} \delta\tilde\alpha^p_j p_{N-j-1/2} 
\end{array}\right) = \hat P \vec{\delta\alpha^p}  
\hspace{5mm}
 \delta D^{(u)} u = 
 \fr{1}{h}\left(\begin{array}{c} 
 \sum\limits_{j=0}^{J} \delta\alpha^u_j u_{j} \\
 0 \\
 \vdots \\
 0 \\
-\sum\limits_{j=0}^{J} \delta\tilde\alpha^u_j u_{N-j} 
\end{array}\right) = \hat U \vec{\delta\alpha^u}  
\eeq
where operators $\hat P$ and $\hat U$ are constructed from the solution  $p$ and $u$ of the original equation.
Their matrices have non-zero elements in the first and in the last lines only: 
\beqr
&&\hat P_{1}=(p_{1/2}, p_{3/2}, \cdots , p_{J+1/2}, \underbrace{0,\cdots , 0}_{J+1\mbox{ times}}),\;
\hat P_{N}=( \underbrace{0,\cdots , 0}_{J+1\mbox{ times}}, p_{N-J-1/2}, \cdots , p_{N-1/2})
\nonumber \\
&&\hat U_{1}=(u_{0}, u_{1}, \cdots , u_{J}, \underbrace{0,\cdots , 0}_{J+1\mbox{ times}}),\;
\hat U_{N-1}=( \underbrace{0,\cdots , 0}_{J+1 \mbox{ times}}, u_{N-J}, u_{N-J+1}, \cdots , u_{N})
\label{hats}
\eeqr
Vectors $\vec{\delta\alpha^p}$ and $\vec{\delta\alpha^u}$ are extracted from matrices $\delta D^{(p)},\delta D^{(u)}$:
\beqr
 \vec{\delta\alpha^p} = ( \delta\alpha^p_0, \delta\alpha^p_1,\delta\alpha^p_2,\ldots,\delta\alpha^p_J,\delta\tilde\alpha^p_{J},\delta\tilde\alpha^p_{J-1},\ldots,\delta\tilde\alpha^p_{0})^t
\nonumber\\
 \vec{\delta\alpha^u} = ( \delta\alpha^u_0, \delta\alpha^u_1,\delta\alpha^u_2,\ldots,\delta\alpha^u_J, \delta\tilde\alpha^u_{J},\delta\tilde\alpha^u_{J-1},\ldots, \delta\tilde\alpha^u_{0})^t
\eeqr
It has to be noted, that operators $\hat P$ and $\hat U$ act from the space of the control variable $\alpha$ to the space of the model's solution $u$ or $p$. Their matrices, consequently,  are  rectangular. Their dimensions are $N\tm 2(J+1)$ and $(N-1)\tm 2(J+1)$ respectively. 

So far, both  $\delta\alpha$ and $(\delta u, \; \delta p)$ are supposed to be small, we neglect their products in \rf{prt} and get  
 \beqr
 \der{\delta u}{t} &=& D^{(p)}\delta p +\hat P \vec{\delta\alpha^p}  \nonumber \\
\der{\delta p}{t} &=&D^{(u)}\delta u +\hat U \vec{\delta\alpha^u}   \label{tlm} 
\eeqr
with the same boundary conditions \rf{bc} for $(\delta u, \; \delta p)$. At initial time both  $\delta u $ and $ \delta p$ are taken to be zero because our study is confined at evolution of a pure perturbation due to boundary scheme.  

The same time stepping as in \rf{timestep} is applied to \rf{tlm}:
\beqr
\fr{\delta u^{n+1}-\delta u^{n-1}}{2\tau} &=& D^{(p)}\delta p^n +\hat P^n \vec{\delta\alpha^p}, \nonumber \\
\fr{\delta p^{n+1}-\delta p^{n-1}}{2\tau} &=&D^{(u)}\delta u^n +\hat U^n \vec{\delta\alpha^u}    \label{timesteptlm}
\eeqr
The first step of the tangent linear model \rf{tlm} is written according to the scheme  \rf{fststep}. Taking into account the zero initial condition 
$\delta u(x,0)=0, \; \delta p(x,0)=0$ we write
\beqr
&&\delta u^{1/2} = \fr{\tau}{2}\hat{P^0} \vec{\delta\alpha^p}, \;\;
\delta p^{1/2} = \fr{\tau}{2}\widehat U^0 \vec{\delta\alpha^u}  \nonumber\\
&&\delta u^{1} = \tau(D^{(p)}\delta p^{1/2} +\hat P^{1/2} \vec{\delta\alpha^p}),
\;\;
\delta p^{1} = \tau(D^{(u)}\delta u^{1/2} +\hat U^{1/2} \vec{\delta\alpha^u})
\label{fsteptlm}
\eeqr
Equation \rf{timesteptlm} can be rewritten  in a  matricial form:
\beq
\left(\begin{array}{c}
\delta u^{n+1}\\
\delta u^{n}\\
\delta\alpha^u\\
\delta p^{n+1}\\
\delta p^{n}\\
\delta\alpha^p\\
\end{array}\right) = 
\left(\begin{array}{cccccc}
0 & I & 0 & 2\tau D^{(p)} & 0 & 2\tau\hat P^n\\
I & 0 & 0 &  0 & 0 & 0\\
0 & 0 & I &  0 & 0 & 0\\
2\tau D^{(u)} & 0 & 2\tau\hat U^n & 0 & I & 0\\
0 & 0 & 0 &  I & 0 & 0\\
0 & 0 & 0 &  0 & 0 & I\\
\end{array}\right)
\left(\begin{array}{c}
\delta u^{n}\\
\delta u^{n-1}\\
\delta\alpha^u\\
\delta p^{n}\\
\delta p^{n-1}\\
\delta\alpha^p\\
\end{array}\right) \label{mattlm}
\eeq
with the first step \rf{fsteptlm}
\beq
\left(\begin{array}{c}
\delta u^{1}\\
\delta u^{0}\\
\delta\alpha^u\\
\delta p^{1}\\
\delta p^{0}\\
\delta\alpha^p\\
\end{array}\right) = 
\left(\begin{array}{cccccc}
 \fr{\tau^2}{2} D^{(p)}\hat U^{0} &  \tau\hat P^{1/2}\\
0 & 0 \\
I & 0 \\
-\tau\hat U^{1/2}&\fr{\tau^2}{2} D^{(u)}\hat P^{0}\\
0 & 0 \\
0 & I\\
\end{array}\right)
\left(\begin{array}{c}
\delta\alpha^u\\
\delta\alpha^p\\
\end{array}\right) \label{matfstlm}
\eeq
To obtain the adjoint model for euclidean scalar product, we introduce adjoint variables 
\beq 
(\phi_u^n,\phi_u^{n+1},\xi_u^n,\phi_p^n,\phi_p^{n+1},\xi_p^n)^t
\eeq
and write backward evolution with  transpose matrices \rf{mattlm} 
\beq
\left(\begin{array}{c}
 \phi_u^{n-1}\\
 \phi_u^{n}\\
 \xi_u^{n-1}\\
 \phi_p^{n-1}\\
 \phi_p^{n}\\
\xi_p^{n-1}
\end{array}\right) = 
\left(\begin{array}{cccccc}
0             & I & 0 & 2\tau (D^{(u)})^* & 0 & 0\\
I             & 0 & 0 & 0             &0 & 0\\
0             & 0 & I & 2\tau(\hat U^n)^* & 0 & 0\\
2\tau (D^{(p)})^* & 0 & 0 &  0            & I & 0\\
0             & 0 & 0 &  I            & 0 & 0\\
2\tau(\hat P^n)^* & 0 & 0 &  0            & 0 & I
\end{array}\right) 
\left(\begin{array}{c}
 \phi_u^{n}\\
 \phi_u^{n+1}\\
\xi^{n}_u\\
 \phi_p^{n}\\
 \phi_p^{n+1}\\
\xi^{n}_p
\end{array}\right) \label{matadj}
\eeq
The last step of the adjoint model is the adjoint of the first step of the tangent model:
\beq
\left(\begin{array}{c}
\xi_u^0\\
\xi_p^0\\
\end{array}\right) = 
\left(\begin{array}{cccccc}
 \fr{\tau^2}{2} (\hat U^{0})^*(D^{(p)})^* & 0&I&  \tau(\hat U^{1/2})^*& 0&0 \\
\tau(\hat P^{1/2})^* &0&0&\fr{\tau^2}{2} (\hat P^{0})^*(D^{(u)})^*&0&I\\
\end{array}\right)
\left(\begin{array}{c}
 \phi_u^{1}\\
 \phi_u^{2}\\
\xi^1_u\\
 \phi_p^{1}\\
 \phi_p^{2}\\
\xi^1_p\\
\end{array}\right)
 \label{matfsadj}
\eeq
where operators $(\hat U^{n})^*, \; (D^{(p)})^*, \; (\hat P^{n})^*, \; (D^{(u)})^*$ are adjoints to \rf{D} and \rf{hats}.

We can see that the right hand side of the  tangent linear model \rf{tlm} is  composed by two  terms:  $D^{(p)}\delta p, \mbox{ or }D^{(u)}\delta u $ and $\hat P \vec{\delta\alpha^p},\mbox{ or } \hat U \vec{\delta\alpha^u} $.  The first one, \rf{D}, is responsible for the evolution of a small perturbation by the model's dynamics, while the second one, \rf{hats},  determines the way how the uncertainty is introduced into the model. The first term is similar for any data assimilation, while the second one is specific to the particular variable under identification. This term is absent when the goal is to identify the initial point because the uncertainty is introduced only once, at the beginning of the model integration. But, when the uncertainty is presented in the approximation of derivatives near the boundary, or some other internal  parameter of the model or of its numerical scheme, the perturbation is introduced at each time step. 

\subsection{Cost function}

To perform variational data assimilation we introduce the following cost function:
\beqr
\costfun(\alpha)&=&\int\limits_0^T \int\limits_0^1 u(\alpha,x,t)-u^{obs}(x,t))^2+(p(\alpha,x,t)-p^{obs}(x,t))^2 dx dt =
\nonumber\\ 
&=&\int\limits_0^T \norme{
\left(\begin{array}{c}
u(\alpha,x,t)-u^{obs}(x,t)\\p(\alpha,x,t)-p^{obs}(x,t)
\end{array}\right)}^2 dt \label{costfn}
\eeqr
where the norm corresponds to Euclidean scalar product
\beq
\norme{\vc{u(x,t)}{p(x,t)}}^2
=\spm{\vc{u(x,t)}{p(x,t)}}{\vc{u(x,t)}{p(x,t)}}=\int\limits_0^1 u^2(x,t) +p^2(x,t) dx \label{norme}
 \eeq
 We suppose we have observations for all variables at any time. For numerical experiments in this paper we shall use the exact solution of the equation \rf{wave} as observations. This will  help  us to see the assimilation procedure and its results in the simplest and clear form. 
 When this technique is applied to more complex model for which the exact solution is not available, we can use either real observations or the model's solution on a finer grid.
 
 To calculate the gradient of the cost function, we  calculate first its variation
 \beqr
 \delta\costfun&=&\costfun(\alpha+\delta\alpha)-\costfun(\alpha)=  
  \nonumber \\
&=& 2\int\limits_0^T \spm{\vc{u(\alpha,x,t)-u^{obs}(x,t)}{p(\alpha,x,t)-p^{obs}(x,t)}}{\vc{ \delta u(x,t)}{\delta p(x,t)}} dt =
 \nonumber \\
 &=& 2\int\limits_0^T \spm{\vc{u(\alpha,x,t)-u^{obs}(x,t)}{p(\alpha,x,t)-p^{obs}(x,t)}}{{\cal T}(t) \vc{ \delta\alpha^u}{\delta\alpha^p}} dt =
  \nonumber \\
  &=& 2\int\limits_0^T \spm{{\cal A}(t) \vc{u(\alpha,x,t)-u^{obs}(x,t)}{p(\alpha,x,t)-p^{obs}(x,t)}}{\vc{ \delta\alpha^u}{\delta\alpha^p}} dt 
\eeqr 
 where ${\cal T}(t) \vc{ \delta\alpha^u}{\delta\alpha^p}$ is the tangent model \rf{matfstlm},\rf{mattlm}  integrated from $t=0$  to $t$ and  ${\cal A}(t)$ is the adjoint model integrated from $t$ to $0$. 
 
 Thus, the gradient of the cost function
 \beq
 \nabla \costfun = 2\int\limits_0^T {\cal A}(t) \vc{u(\alpha,x,t)-u^{obs}(x,t)}{p(\alpha,x,t)-p^{obs}(x,t)} dt 
 \label{grad}
 \eeq
 is obtained as the sum of the adjoint model integrations. Each   
integration of the adjoint model  starts from multiplication of the matrix \rf{matadj} by the vector
$$
\left(\begin{array}{c}
 u(\alpha,x,t)-u^{obs}(x,t)\\
 0\\
0\\
p(\alpha,x,t)-p^{obs}(x,t)\\
 0\\
0
\end{array}\right)
$$
and followed by subsequent multiplications by matrices \rf{matadj} taken at corresponding time. This product is finally multiplied by the matrix \rf{matfsadj} to get the vector $\left(\begin{array}{c}
\xi_u^0\\
\xi_p^0\\
\end{array}\right)$ which represents the gradient of the cost function.

This gradient is used in the minimization procedure that is implemented in order  to find the minimum
\beq
\costfun(\bar\alpha) = \min_{\alpha} \costfun(\alpha)
\eeq
Coefficients $\bar\alpha$  are considered as coefficients realizing  optimal discretization of the model's operators in the boundary regions. 

The  minimization procedure used here was developed by Jean Charles Gilbert and  Claude Lemarechal, INRIA \cite{lemarechal}.  The procedure uses the limited memory quasi-Newton method.

\section{Results of assimilation}

Exact solution of the equation \rf{wave} can easily be found by the method  of variables separation. We look for  solutions in a special  form 
$u(t,x)=a(t)b(x)$. A consequence is that 
$$\fr{a"}{a}=\fr{b"}{b}=-\lambda.$$
The value of $\lambda$ is determined so that there exists a non-trivial solution of the boundary-value problem
$$b"+\lambda b =0, \hspace{1cm} b(0)=b(1)=0$$
Values of $\lambda$ are all positive, and the solutions are trigonometric functions. A solution that satisfies square-integrable initial conditions \rf{ic} for  $u$ and $p$ can be obtained from expansion of these functions in the appropriate trigonometric series. 

\subsection{One trigonometric mode}

We shall analyze first the behavior of one trigonometric mode of the solution and further proceed with the analysis of more complex functions. 

Let us define the initial point for $u$ and $p$ in \rf{wave} as
\beq
u(x,0)=sin(k\pi x) \;\; p(x,0)=cos(k\pi  x)
\eeq
The solution of \rf{wave} determined by $\lambda=k^2\pi^2$ is
\beq   
 u_{exact}(x,t)=-\sqrt{2} \sin(k\pi t-\pi/4) \sin(k\pi x), \;\; p_{exact}(x,t)=\sqrt{2} \cos(k\pi t-\pi/4) \cos(k\pi x) \label{exact}
 \eeq
 The solution \rf{exact} will be used as artificial ``observations" to be assimilated into the discretized  wave equation. The use  of these  data allows us to   work with errors of numerical  schemes only,  avoiding all additional errors that may be present due to inexact data. 
 
Two  numerical approximations are used for discretization of spatial derivatives  in all internal points of the interval. Both discretizations are performed by formula \rf{intrnlsch}, but one of them is of second order of accuracy with coefficients  $ a_j=(0,-1,1,0)$ for $ j=(-1,0,1,2)$ and the other one is of fourth order with $ a_j=\fr{1}{24}(1,-27,27,-1)$. The simplest second order scheme on the boundary was used in both cases.  That means both $\alpha^u_j$ and $\alpha^p_j$ in \rf{bndsch} were chosen to provide classical approximation of derivatives in points adjacent to boundary:  
\beq
\biggl(\der{p}{x}\biggr)_1=\fr{p_{3/2}-p_{1/2}}{h}, \;\;
\biggl(\der{u}{x}\biggr)_{1/2}=\fr{u_1-u_0}{h}  \label{classic}
\eeq

In order to see precisions of these schemes   we calculate the difference between the  numerical solution $u(x,t),p(x,t)$ and the exact one $u_{exact}(x,t),p_{exact}(x,t)$ and plot its norm 
\beq
\xi(t)=\int\limits_0^1\biggl((u(x,t)-u_{exact}(x,t))^2+(p(x,t)-p_{exact}(x,t))^2\biggr) dx.
\label{xi}
\eeq
Numerical solutions are obtained with $k=3$, $h=\fr{1}{30}$ and $\tau=\fr{1}{120}$. 

It is well known that   the principal error of classical (with   approximations of derivatives near the boundary realized  by \rf{classic}) solutions for both second and fourth order schemes consists  in the wrong wave speed. 
Numerical solution of \rf{wave} is also composed of trigonometric functions of the same amplitude but they oscillate with wrong frequency. The second order solution oscillates  a little slower than the exact one, and the fourth order oscillates a little faster.

In \rfg{error-orig-xt}A and \rfg{error-orig-xt}B we see that the difference between exact and numerical solutions oscillates with the frequency $3\pi$ but have a growing amplitude.  The velocity error is lower when the fourth order approximation is used, that's why the amplitude of the difference in  \rfg{error-orig-xt}B is lower than in  \rfg{error-orig-xt}A. 

\figureleft{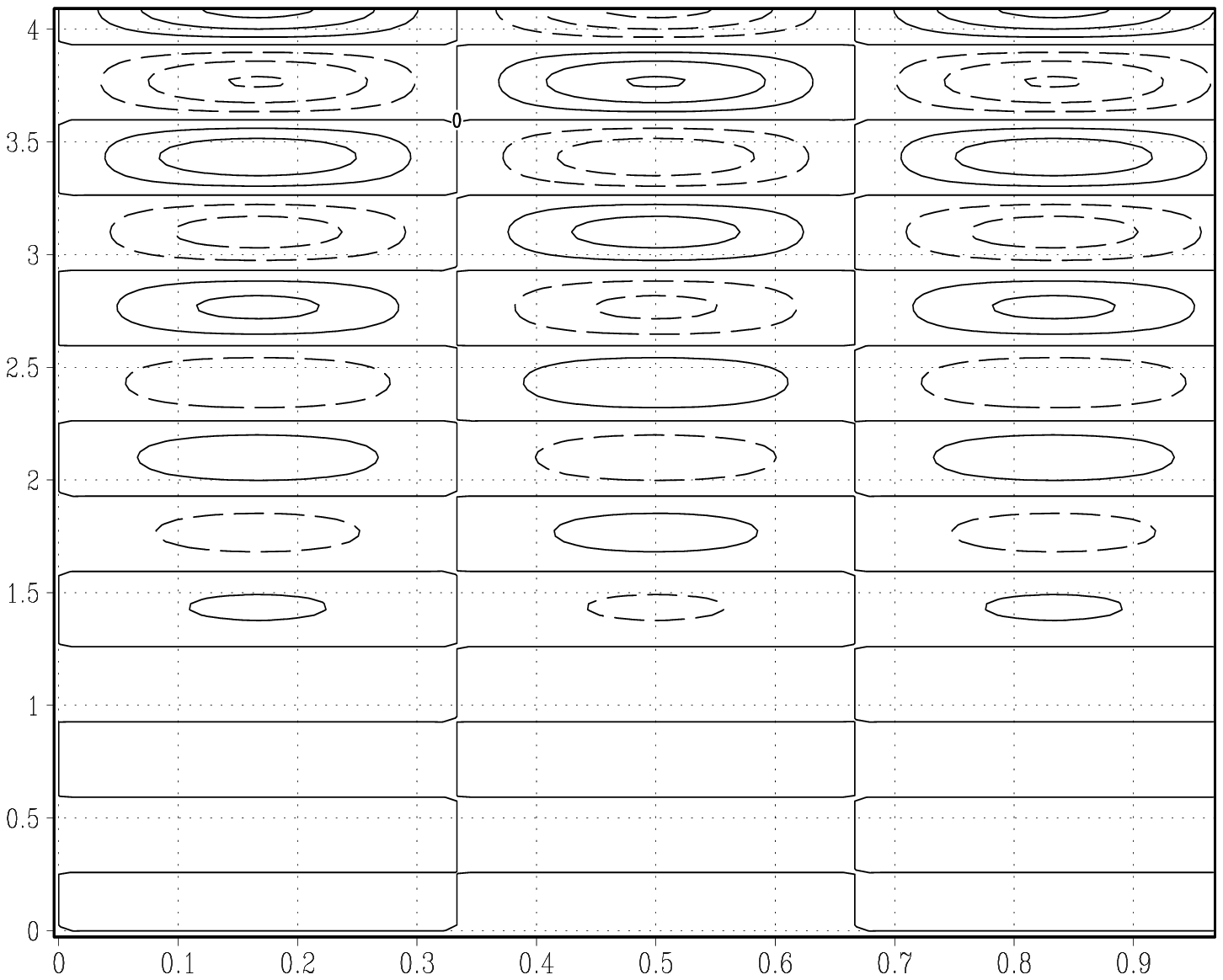}{ $x-t$ diagram of the error of the classical second order scheme. Contours from -0.2 to 0.2 with interval 0.05.  } {error-orig-xt} 
\figureright{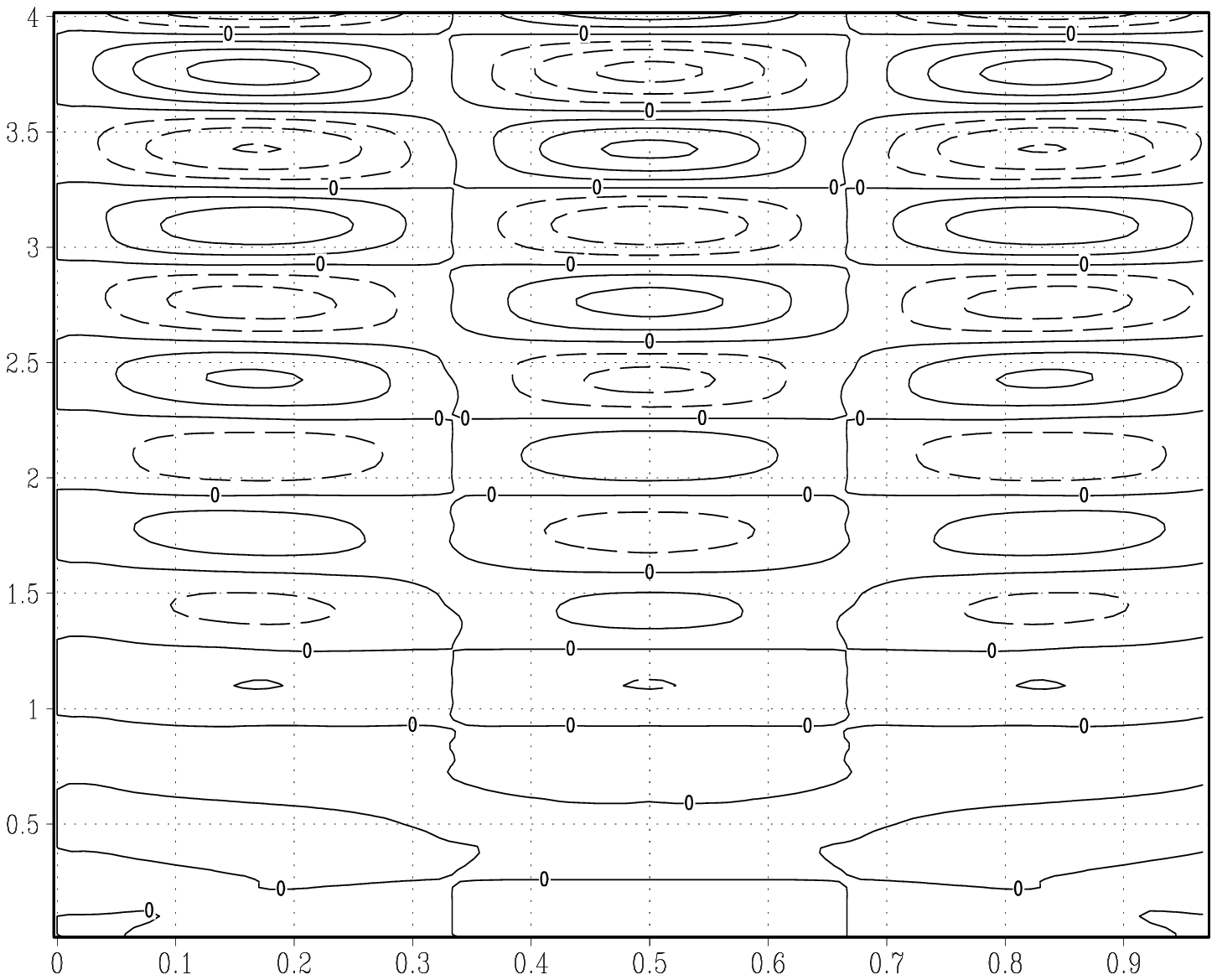}{ $x-t$ diagram of the error of the classical fourth order scheme. Contours from -0.04 to 0.04 with interval 0.01. }

If we look at figures \rfg{errors-int}A and 2B, we see the same phenomenon. The solid line in \rfg{errors-int}A, that  represents the norm of the difference between the exact solution and its second order numerical approximation, grows first up to value of 120 at time $t=108.3$ time units.  After that, the norm decreases to 0 at  time $t=215.9$ and restarts to grow. The fourth order approximation exhibits a similar behavior, the norm  also grows up to value 120, but it reaches its maximum  and the following   zero at  $t=491.1$ and   $t=982.2$ time units respectively. These moments of time, being beyond the picture window, are   not shown.  The speed error in the second order approximation results that at time $t=215.9$ numerical solution is exactly one wave period later than the exact one, and the difference between them vanishes. So far, the speed error is lower for the fourth order approximation, moments of the maximal and vanishing norm in \rfg{errors-int}B are reached later.  

\figureleft{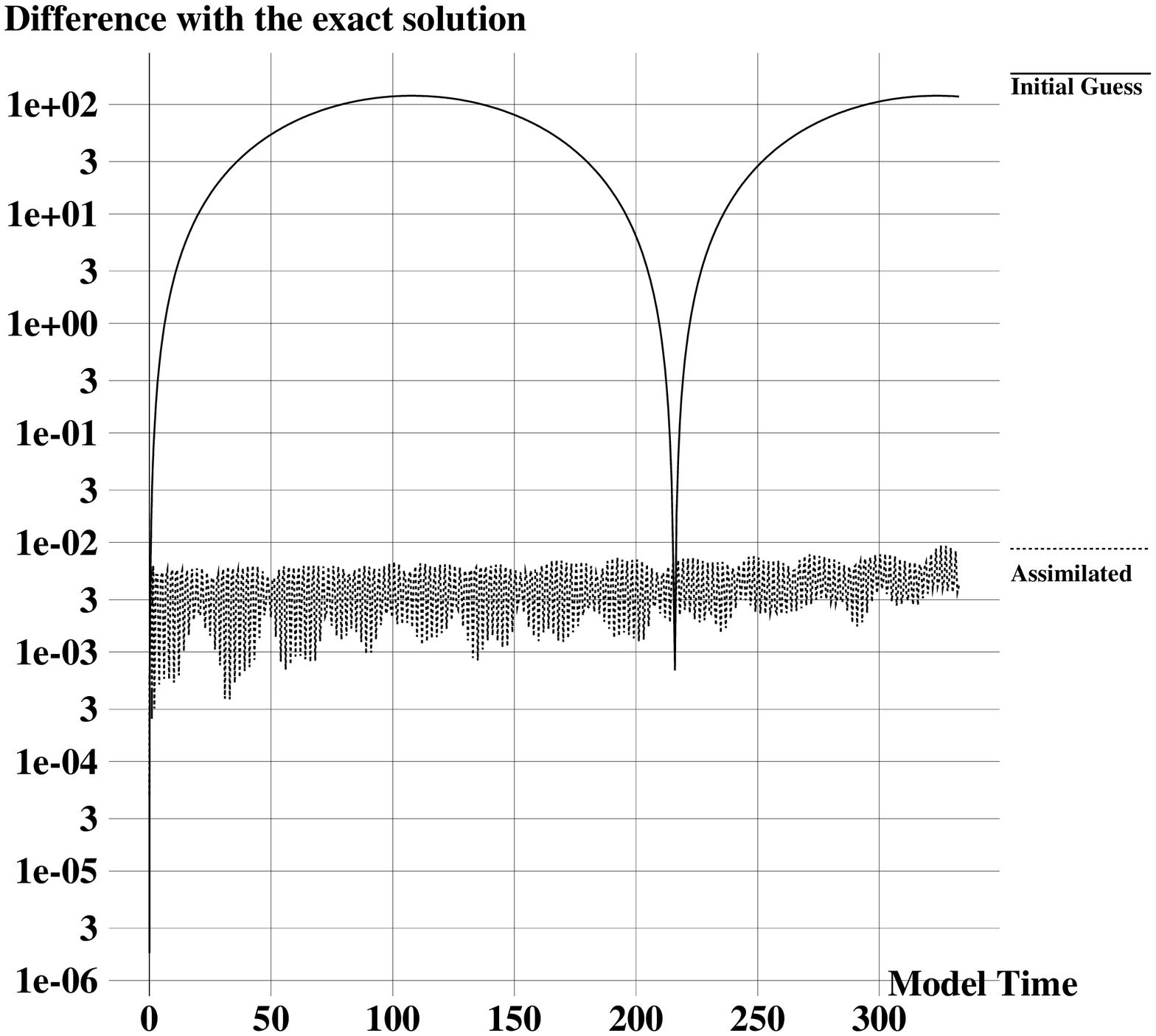}{ Error $\xi(t)$ of the second order scheme: Classical -- solid line, with assimilated boundary -- dashed line.  } {errors-int} 
\figureright{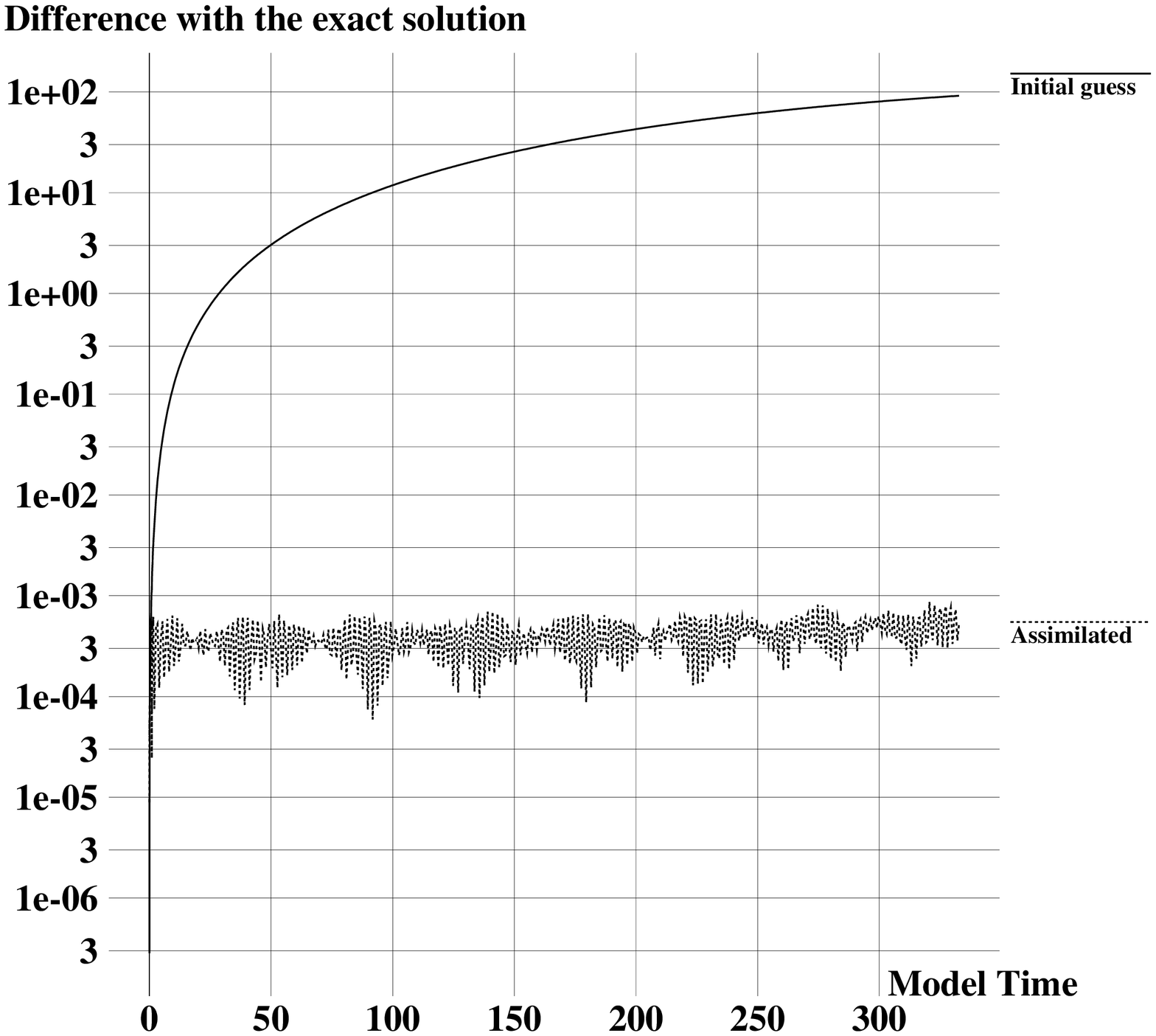}{Error  $\xi(t)$ of the fourth order scheme: Classical -- solid line, with assimilated boundary -- dashed line. }

Thus, it was illustrated that  the principal error of numerical approximation consists in the wrong wave speed. 
Indeed,  if we apply numerical approximations to trigonometric functions,  we can calculate the error in the wave velocities.  We substitute trigonometric solutions for $u$ and $p$ in the second order scheme,
\beqr
u(x,t)&=&\sin(kx)\sin(kt)=\sin(ikh)\sin(nk\tau) \nonumber\\
p(x,t)&=&-\cos(kx)\cos(kt)=-\cos(ikh)\cos(nk\tau)\nonumber
\eeqr
 we get
\beqr
\biggl(\der{p}{x}\biggr)_i^n&=&\fr{p_{i+1/2}^n-p_{i-1/2}^n}{h}=
\cos(nk\tau)\fr{\cos((i+1/2)kh)-\cos((i-1/2)kh)}{h}=
\nonumber \\
&=&\fr{2\sin(\fr{kh}{2})\sin(ikh)\cos(ik\tau)}{h} \nonumber \\
\biggl(\der{u}{t}\biggr)^n_i&=&\fr{u_i^{n+1}-u_i^{n-1}}{2\tau}=
\sin(ikh)\fr{\sin((n+1)k\tau)-\sin((n-1)k\tau)}{2\tau}=
\nonumber \\
&=&\fr{\sin(k\tau)\sin(ikh)\cos(ik\tau)}{\tau} \nonumber
\eeqr
Thus, the first equation in \rf{wave} is approximated by
\beqr
\biggl(\der{u}{t}\biggr)^n_i - \biggl(\der{p}{x}\biggr)_i^n &=& 
\biggl(\fr{\sin(k\tau)}{\tau}-\fr{2\sin(kh/2)}{h}\biggr) \sin(ikh)\cos(ik\tau)=
\nonumber\\
&=&
\fr{h \sin(k\tau)-2\tau \sin(kh/2)}{2\tau \sin(kh/2)}\biggl(\der{p}{x}\biggr)_i^n
\eeqr
Similar substitutions for $u$ and $p$ in the second equation give us the approximation of the system
\beqr
\biggl(\der{u}{t}\biggr)^n_i - \beta_2 \biggl(\der{p}{x}\biggr)_i^n = 0 \nonumber \\
\biggl(\der{p}{t}\biggr)^n_i - \beta_2 \biggl(\der{u}{x}\biggr)_i^n = 0\label{cngvlc} 
\eeqr
with
\beq
\beta_2=\fr{h \sin(k\tau)}{2\tau \sin(kh/2)} \label{beta2}
\eeq
Thus, we see that numerical wave velocity is equal to $\beta_2$ rather than to one. 

If we perform similar manipulations with the fourth order spatial discretization, i.e. approximation of all spatial derivatives by \rf{intrnlsch} with  stencil $ a_j=\fr{1}{24}(1,-27,27,-1)$, we get the velocity error
\beq
\beta_4=\fr{12h \sin(k\tau)}{27\tau \sin(kh/2)-\tau \sin(3kh/2)} \label{beta4}
\eeq
In \rfg{vlcerr} we can see the form of speed errors $\beta_2-1$ and $\beta_4-1$  for three values of $k$. Horizontal axis is marked in values of $\fr{\tau}{h}$.

\figurecent{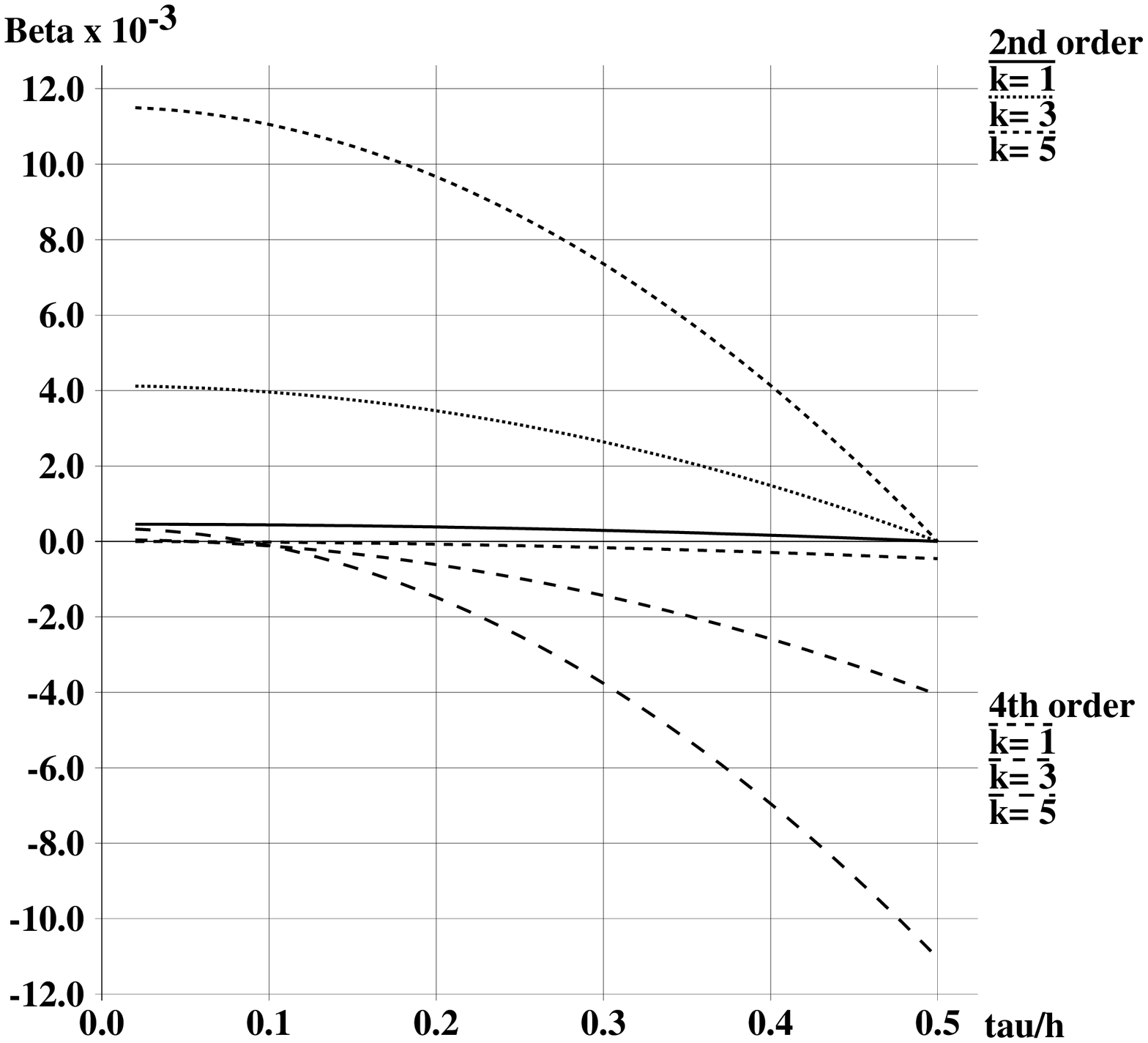}{ Wave speed error ($\beta-1$) for second and fourth order schemes as functions of $\tau/h$. } {vlcerr}

 We see that  using second order scheme, we can simulate the exact solution. Indeed, when $h=2\tau$ the velocity of numerically approximated wave is exactly equal to the velocity of the theoretical solution for any wave-number $k$.  Using any lower $\tau$ we must assume the error in the waves velocity. 
 
 On the other hand, it is impossible to calculate an exact solution with a fourth order scheme. The value of $\beta_4-1$  vanishes in different points $\tau/h$ for different $k$. The only conclusion we can make is the ratio $\tau/h$ must either be small for this scheme, or some higher order time stepping should be used. 
 
For the given parameters ($k=3,\;  h=\fr{1}{30}$ and $\tau=\fr{1}{120}$) errors in the wave velocity can be calculated by \rf{beta2} and \rf{beta4}:
$$\beta_2=3.09\tm 10^{-3} \;\; \beta_4=-9.82\tm 10^{-4}$$
These velocity errors determine the time when the numerical wave will be one period shifted with respect to the exact wave:
$ T=\fr{\mbox{wave period}}{\beta}=\fr{2}{k\beta}$. For the second order scheme with $k=3$ this time $T$ is equal to  $215.6$ time units that corresponds well to numerically obtained $215.9$.

So, knowing errors produced by  numerical schemes with  chosen parameters, we shall perform the  assimilation of the exact solution  in order to see how these errors can be corrected by  the optimal boundary discretization. 

We perform the data assimilation minimizing the cost function \rf{costfn} assuming that  the approximation of boundary derivatives is composed by two terms only ($J$ in \rf{bndsch} is equal to 1) and we get a numerical solution with no error in wave velocity. The norm  \rf{xi}  of  the difference between the exact solution and its optimal numerical approximation  (lower lines in \rfg{errors-int}A and B) oscillate around $3\tm 10^{-3}$ and $3\tm 10^{-4}$ respectively. $X-t$ plots of the difference $u(x,t)-u_{exact}(x,t)$ presented in \rfg{error-assim-xt}A and  \rfg{error-assim-xt}B show very similar behavior of the error.  The difference is composed of small moving waves that propagate back and forth between the boundary and the middle of the interval  for both the second  and  the fourth order schemes.  The amplitude of these waves is small comparing to errors of the classical scheme and, that is more important, remain small during any integration time. This fact can be seen in \rfg{errors-int}. Despite the data were assimilated during 6 time units only ($T$ in \rf{costfn} is equal to 6),  boundary approximation of derivatives has been sufficiently well  identified to satisfy the model during any long integration, 300 time units and more.

\figureleft{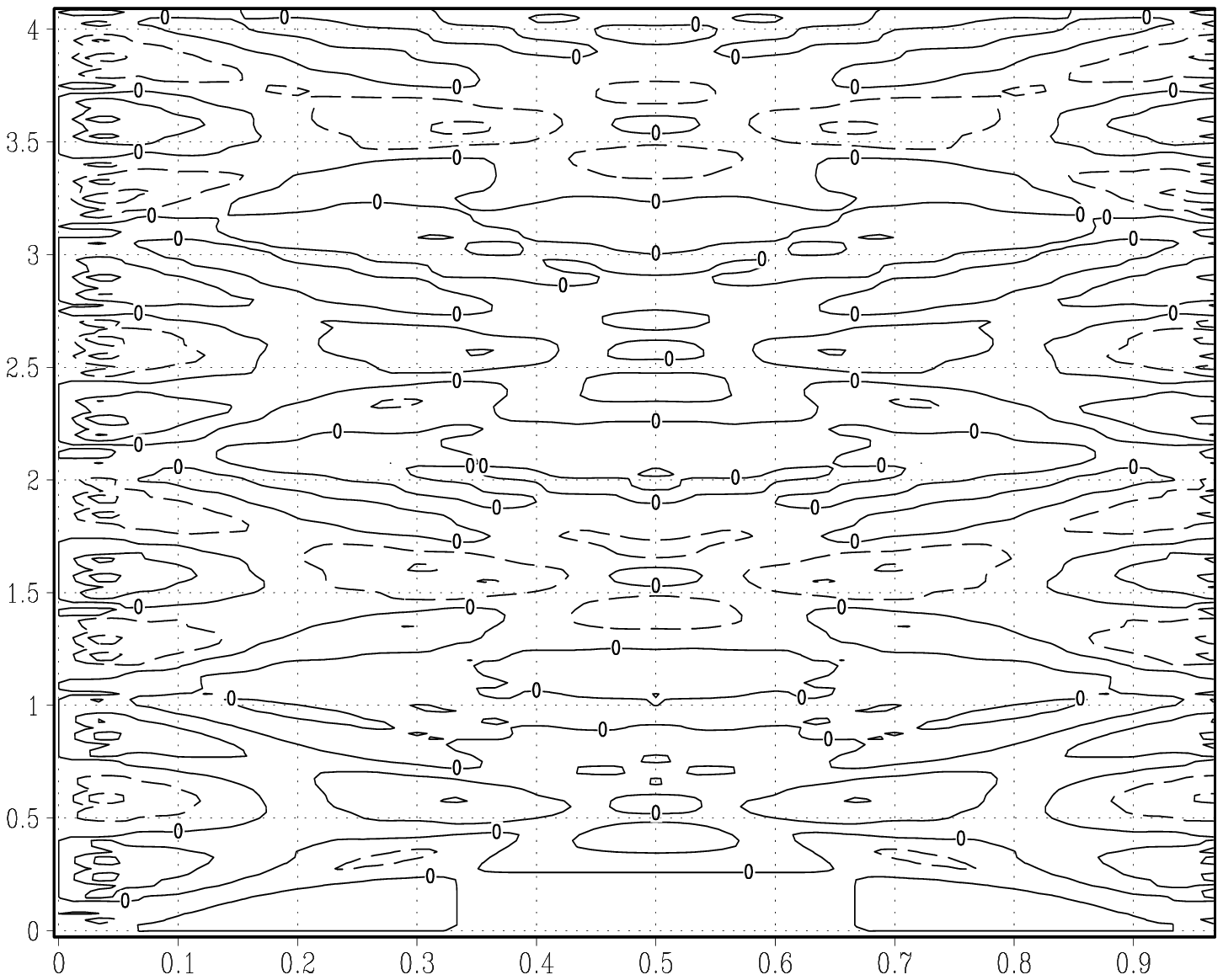}{ $x-t$ diagram of the error of the modified second order scheme. Contours from -0.03 to 0.03 with interval 0.01.   } {error-assim-xt} 
\figureright{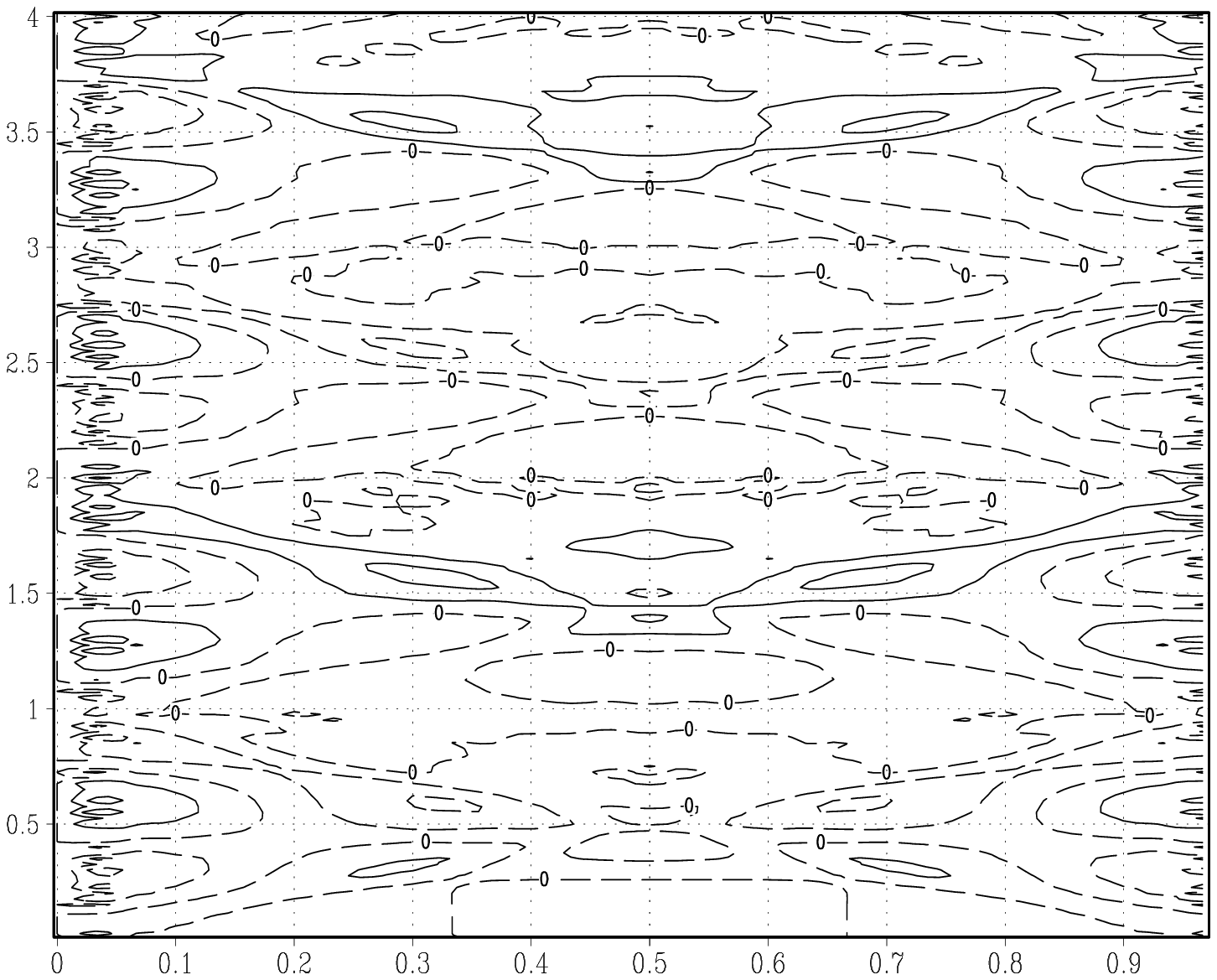}{$x-t$ diagram of the error of the modified  fourth order scheme. Contours from -0.01 to 0.01 with interval 0.003.   }

The choice of an optimal  assimilation window (the time interval $T$ during which the assimilation is performed) is obvious for this simple problem. Of course, $T$ must not be too small. It must cover at least several wave periods in order to provide necessary information about errors in wave velocities. On the other hand, too long  $T$ is not optimal, because the assimilation  over a longer interval  is less  efficient. First, we do not need too much data to assimilate because of   the simplicity of the model.  And second, too long $T$ reduces computational efficiency of the method because of the necessity to run the model for a longer time in each iteration.

Thus, assimilating the exact solution of the equation, we can construct an optimal approximation of boundary derivatives and obtain a rather accurate  model which error is sufficiently small. However, boundary derivatives obtained in this procedure are strange from the point of view of approximation. 

When the  second order approximation is used for derivatives in all internal points of the interval,  the optimal discretization  near the boundary obtainded by data assimilation has a form 
\beq
\biggl(\der{u}{x}\biggr)_{1/2}=1.048\fr{u_1-u_0}{h}, \quad
\biggl(\der{p}{x}\biggr)_{1}=\fr{3.014 p_{3/2}-2.828 p_{1/2}}{h} \label{1.048}
\eeq
First of all, these formulas do not approximate a derivative. The first one approximates the derivative multiplied by 1.048, the Taylor expansion of the second one  has a form
$$
0.18 \fr{p_1}{h}+2.92 \biggl( \der{p}{x}\biggr)_1 +0.023 h \biggl( \fr{\partial^2 p}{\partial x^2}\biggr)_1
+0.12 h^2 \biggl(  \fr{\partial^3 p}{\partial x^3}\biggr)_1 +O(h^3)
$$

Neither expression for $\der{u}{x}$, nor for $\der{p}{x}$ has any reasonable order of approximation. The first one is of 0 order, the second is of -1 order. Moreover, while we get always the same formula for  $\der{u}{x}$,  approximation of the derivative of $p$ varies in different assimilation experiments. Assimilations performed with different assimilation windows, for example, result in different coefficients for $ \der{p}{x}$. In fact, any combination $\alpha^p_0\; , \alpha^p_1$ in \rf{bndsch} may be found as the result of assimilation under condition 
\beq\alpha^p_1=-1.104 \alpha^p_0 -0.107.\label{p-line}\eeq

 This linear relationship has been obtained experimentally performing assimilations with all assimilation windows in range from 600 to 2400  time steps (with the  time step equal to $1/120$ of the  time unit). Resulting couples $\alpha^p_0\; , \alpha^p_1$ presented in \rfg{alphap} are positioned on  a straight line with values $\alpha^p_0$ varying from -1.5 to -5. 

\figurecent{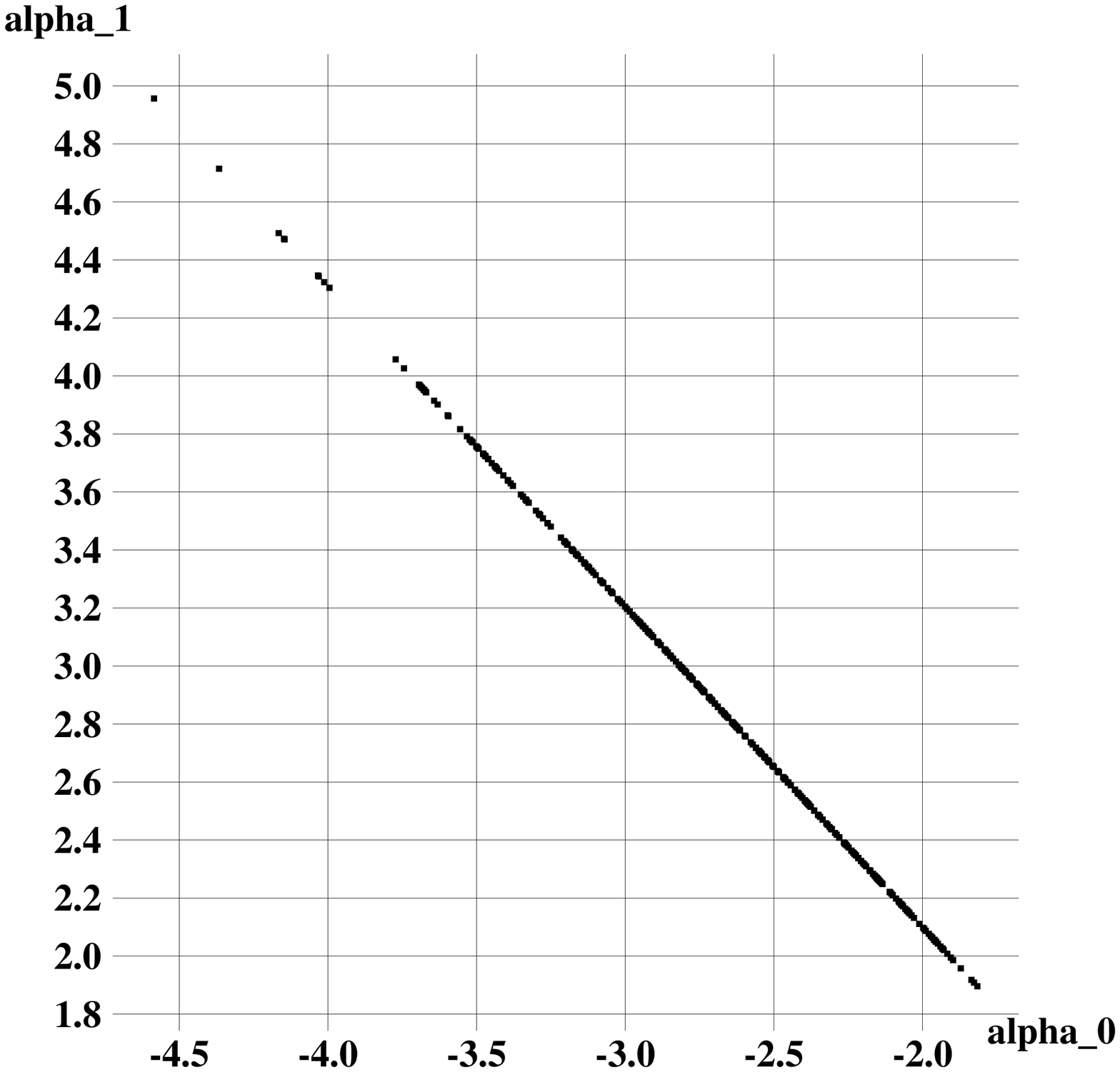}{ Scatter diagram of $\alpha^p_0\; , \alpha^p_1$ obtained with different assimilation windows $T$ in range from 5 to 20 time units. } {alphap}

To explain these unusual  approximations of the derivatives, we address first the $u$ derivative, that is always approximated by $1.048\fr{u_1-u_0}{h}$. We know, the principal error of the classical scheme consists in wrong wave velocity. The data assimilation and control of the boundary derivatives can not modify numerical wave velocity. The only way for this control to get a better solution consists in modifying  the length of the interval. A numerical wave with wrong velocity will propagate on the interval with wrong length. But the length of the interval is adapted by data assimilation in order  to ensure the  wave with  numerical velocity propagates the modified interval in the same  time that the exact wave propagates the exact interval.  So far,  the control can not correct the error in the wave velocity, it commits another error in length in order to compensate  the first one. 

As we have seen, the coefficient 1 in front of spatial derivatives in \rf{wave} has been replaced by $\beta$ in \rf{cngvlc}. Theoretical wave speed $c_{exact}=1$ has, consequently, been replaced by numerical speed $c_{num}=\beta$. The length of the interval $L_{exact}=1$ should also be modified  to satisfy
\beq
\fr{L_{exact}}{c_{num}}=\fr{L_{modified}}{c_{exact}} \Longrightarrow 
L_{modified}= L_{exact}\fr{c_{exact}}{c_{num}}=\fr{L_{exact}}{\beta}
\eeq
However, the control can not  modify all grid cells of the interval uniformly.   It can act near boundaries only and can modify the length of  cells just  adjacent to boundary points. Hence, only two grid cells, one on the left and one on the right of the interval,  can be modified. The modified interval, hence, becomes composed by $N-2$ cells of length $h=\fr{1}{N}$ and two boundary cells of length
\beq
2h_{modified}+(N-2)h=\fr{hN}{\beta}\; \Longrightarrow 
\fr{h_{modified}}{h}= 1-\fr{N}{2}\fr{\beta-1}{\beta} \label{hmod}
\eeq
For given parameters  (N=30) $\beta_2=1+3.09\tm 10^{-3}$   the boundary cells must be reduced to $h_{modified}=(1-0.046)h$. Consequently, the derivative $\biggl(\der{u}{x}\biggr)_{1/2}$ must  be calculated over modified cell
$$\biggl(\der{u}{x}\biggr)_{1/2} = \fr{u_1-u_0}{h_{modified}}= \fr{1}{1-0.046}\fr{u_1-u_0}{h}= 1.048\fr{u_1-u_0}{h}$$
This is exactly the coefficient obtained in the data assimilation for the derivative of $u$ \rf{1.048}. 

So, we can state that it is reasonable to obtain wrong approximation of derivatives near boundaries as a result of data assimilation. This error compensates the error of the wave speed. 

As for derivatives of $p$, they must also be modified. The only difference with $u$ consists in fact that $\biggl(\der{p}{x}\biggr)_{1}$ is calculated over two half of cells: one half of the first cell (adjacent to boundary point), and one half of the second one, next to the first. Hence, only one half of the modified cell participates in the derivative of $p$ and its modification is
\beq
\biggl(\der{p}{x}\biggr)_{1}=\fr{p_{3/2}-p_{1/2}}{h_{modified}/2+h/2} = \fr{p_{3/2}-p_{1/2}}{h}\fr{2h}{h_{modified}+h}
\eeq
In this experiment we should have obtained $\biggl(\der{p}{x}\biggr)_{1}= 1.023\fr{p_{3/2}-p_{1/2}}{h}$.  

And indeed, the couple $\alpha^p_0=-1.023\; , \alpha^p_1=1.023$ belongs to the set \rf{p-line}. This is the only point on this line where $\alpha^p_1+ \alpha^p_0=0$ and the derivative is approximated with zero order rather than minus first order.  

Non uniqueness of optimal $\alpha^p_1$ and  $\alpha^p_0$  can be explained if we take into account that $p$ has also a form of cosine of $3\pi x$. Hence, at any time $p_{1/2}=A(t)\cos(3\pi h/2)$ and $ p_{3/2}=A(t)\cos(9\pi h/2)$ with some $A$ depending on time.  Their linear combination $\alpha^p_1 p_{1/2}+ \alpha^p_0 p_{3/2}$ can vanish if \beq\alpha^p_1=-\fr{\alpha^p_0}{4\cos^2(k\pi h/2)-3}.\label{zerocomb}\eeq
 Consequently, all couples  $\alpha^p_1\;, \alpha^p_0$ belonging to the  line that passes by the point 
$\alpha^p_0=-1.023\; , \alpha^p_1=1.023$ with tangent $-\fr{1}{4\cos^2(3\pi h/2)-3}= -1.108$
produce the same derivative. This line  coincides withing accuracy of computation  with the set \rf{p-line} obtained numerically.  Any point on this line gives coefficients $\alpha_p$ that theoretically  provide the same value of the derivative and the same value of the cost function.   This line forms the kernel of the Hessian of the cost function. 

Numerical approximation of the solution is slightly different from cosine and numerical approximations of the derivative obtained with different coefficients $\alpha^p$ from the kernel are not exactly the same. The assimilation chooses the best fitting  point in the kernel for particular experiment that provides slightly lower value of the cost function. The choice of this point depends on particular  parameters of the experiment such  as assimilation window. That's why we get different pairs 
$\alpha^p_0, \alpha^p_1$ in different experiments. All these pairs are in the kernel of the Hessian, they provide almost the same cost function values, but each  of them corresponds better to one  particular window.  
If we are interested in optimal boundary scheme for the whole model rather than  in the best fitting point for a given assimilation window, we may  define another criterium of choice and impose this criterium in the cost function.  One choice, usually assumed in data assimilation, requires that optimal point must be  situated not far from the initial guess. 
However, adding this requirement would not allow us to  choose one point in the kernel. The requirement of low distance from the start would draw the optimal point out of the kernel because, as we have seen above, the initial guess point is not situated in the kernel.   

  Instead of imposing low distance from the starting point of minimization, we prefer to require the term in the Taylor expansion with the order minus one to be equal to zero. This implies the sum $ \sum\limits_{j=0}^{J} \alpha_j =0$ must vanish.
For this purpose we add the term
\beq
R=\eta (\sum\limits_{j=0}^{J} \alpha_j )^2 \label{dumpcnst}
\eeq
to the cost function \rf{costfn} and appropriately modify its gradient \rf{grad} adding the term
\beq
\nabla R= 2\eta  \sum\limits_{j=0}^{J} \alpha_j.
\eeq  
Imposing sufficiently large  weight $\eta$ we get the only  approximation of $p$ derivative for any assimilation window. The derivative is approximated by   $\alpha^p_0=-1.023\; , \alpha^p_1=1.023$ that ensures vanishing first term in the Taylor development. 

Modification of the cost function by \rf{dumpcnst} has a very small  influence on the final value of the cost function because this modification determines the choice of the particular point in the kernel of the Hessian. 

Finally, we note that there is no significant difference in the final value of the cost function in experiments with different $J$ in \rf{bndsch}. Several experiments have been carried out with 2, 3 and 5 controlled coefficients $\alpha$, but the minimization procedure has converged always to the same value. 
Obviously, two control coefficients  are already sufficient in this simplest case.  Adding supplementary $\alpha$  just increases the kernel dimension with no influence on the cost function. 

\subsection{Two trigonometric modes}

When initial conditions of the model \rf{wave} are more complex than one trigonometric mode, the exact solution of the wave equation is a linear superposition of exact solutions corresponding to each trigonometric mode of the Fourier development of initial conditions. Each mode has it's own frequency and propagates with it's own velocity. 

 Numerical solution for each Fourier mode commits an error in the wave velocity. But, as it  has been discussed  above, this error is different for different modes because it depends explicitly on the wavenumber $k$ \rf{beta2}, \rf{beta4}. Consequently, in presence of multiple Fourier modes, the interval length must be modified in order to correct different errors in wave's velocities simultaneously. 

We consider first a superposition of just two waves  with $k=2$ and $k=5$. We see from the equation \rf{hmod} that to compensate the error in the wave velocity for the wave with $k=2\pi$, the control must modify the length of the boundary cell by $\fr{h_{modified}}{h}=1-0.020 $ and the coefficient in front of the approximation of the derivative of $u$ at point $1/2$ must be $1.021$.  In the same time, the velocity error for the wave with $k=5\pi$ is compensated when $\fr{h_{modified}}{h}=1-0.128 $ and the coefficient in front of the  derivative $1.142$. 

Performing   experiments with both wavenumbers $k=2$ and $k=5$ separately and with their superposition, we see in \rfg{2wv}A that the data assimilation procedure  is able to compensate the error in wave velocity in all three cases.  The cost function of the model with original coefficients shows wrong velocities of numerical waves in all three experiments, but the model's solution with optimal coefficients is much closer to the exact one. We see the  cost function values as low as $3\tm 10^{-4}$ for the wave with $k=2$ and $ 10^{-1}$ for the wave with $k=5$. The line that corresponds to  the cost function in the experiment with two waves superposed is indistinguishable from the line corresponding to the experiment with $k=5$.  They oscillate both  around $ \costfun=10^{-1}$. That means the residual error of  assimilation of the superposition of two waves is close to the biggest error of assimilation of each particular wave. 

In order to  analyze the expression that is used to calculate the derivative of $u$ near boundaries in \rfg{2wv}B, we perform a set of  assimilations with all assimilation windows in range from 600 to 2400  time steps (with the  time step equal to $1/120$ of the  time unit) for  all three types of initial conditions of the model, i.e. one wave with either $k=2\pi$ or $k=5\pi$ and both of them. When $k=2\pi$ we get always the same resulting couples $\alpha^u_0=-1.021\; , \alpha^u_1=1.021$ as expected. Coefficients $\alpha^u_0\; , \alpha^u_1$ in the experiment with $k=5\pi$ are also all positioned near the theoretical value $\pm 1.142$, but not as concentrated as in the experiment with $k=2\pi$. Values in this experiment are distributed in the interval from 1.138 to 1.144. Obviously, the wave length of the wave with $k=5\pi$ is too short to be well reproduced by a 30 points resolution  grid. This coarse resolution adds numerical noise in the solution and leads to the dependence of the assimilation result on the window.  

Optimal coefficients $\alpha^u_0\; , \alpha^u_1$ in the experiment with two waves are situated in the middle of the figure \rfg{2wv}B.  We can note two particularities. First, their distribution is even more dispersive than with  $k=5\pi$: they occupy the interval from 1.07 to 1.09. And second, expressions for $u$ derivative near the left and near the right boundary are no longer the same. One can see in \rfg{2wv}B the set of coefficients  $\alpha^u_0\; , \alpha^u_1$ in this experiment is splitted into two subsets with a gap between them. 

\figureleft{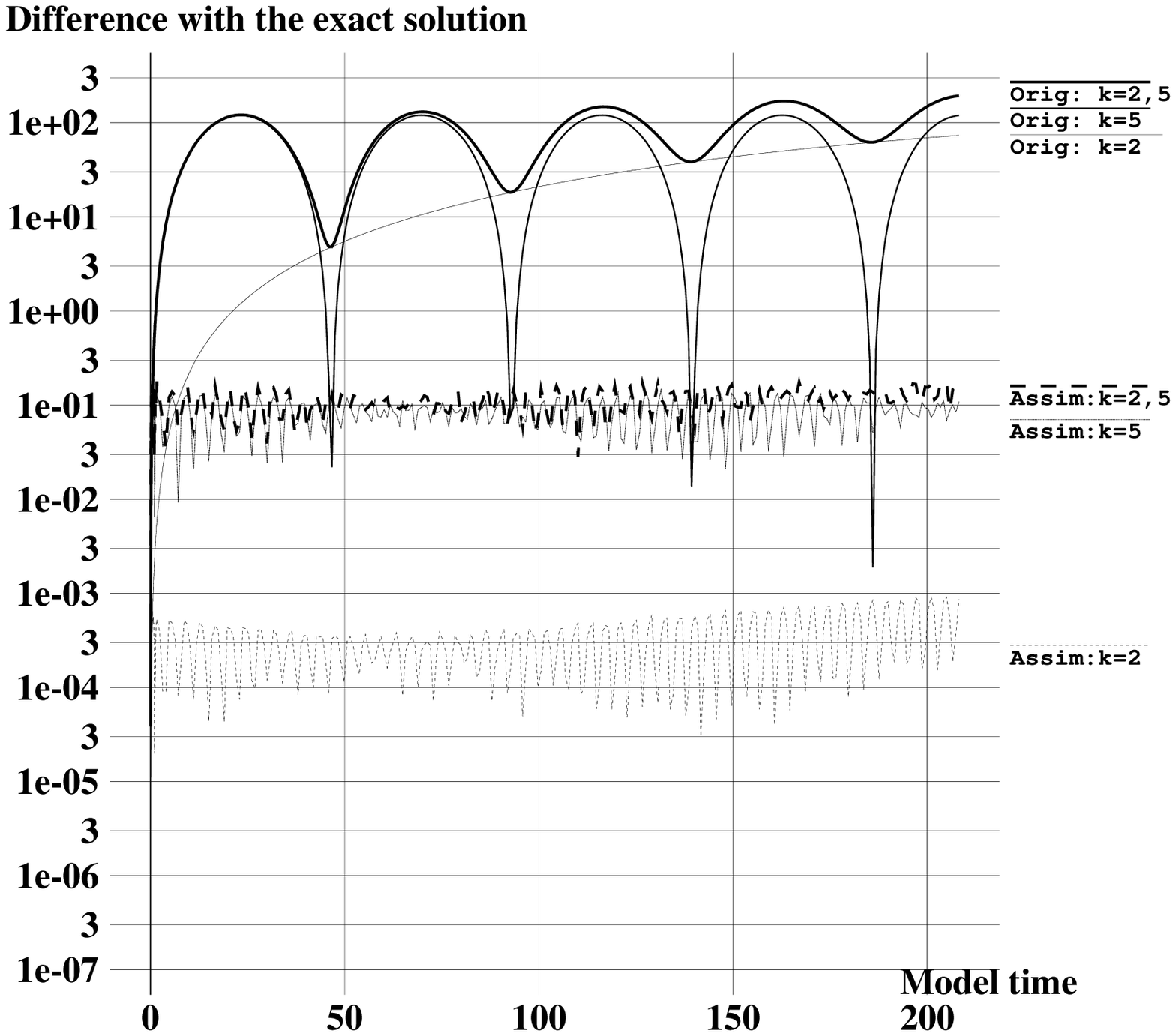}{ Cost function of numerical solutions  for modes with $k=2\pi$, $k=5\pi$ and their superposition. Solutions with the original boundary scheme  are plotted with solid lines, with identified schemes -- by dashed lines.  } {2wv}
\figureright{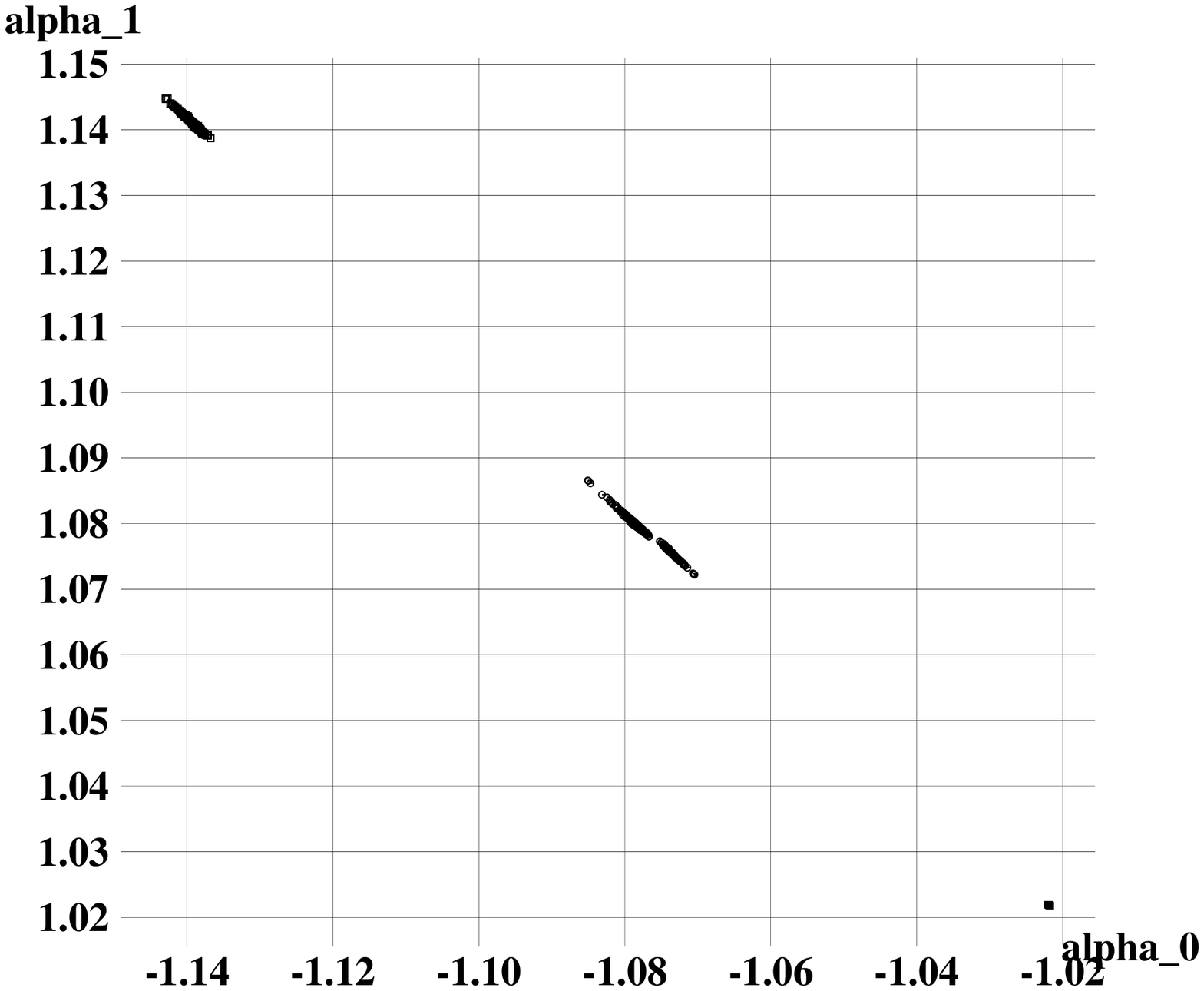}{ Optimal coefficients for $\der{u}{x}$ approximation  for modes with $k=2\pi$ (lower right corner), $k=5\pi$ (upper left corner) and their superposition (center).  }

Coefficients of expressions for $p$ derivative in the experiments with two waves (not shown) possess also a kernel that form  the line situated between lines obtained in experiments with single waves. 

\subsection{Other functions }

If we consider an arbitrary functions as initial conditions of the wave equation, we have all admissible Fourier modes in the solution. In order to see the action of the control in this situation we perform the data assimilation for the model with initial conditions prescribed as 
\beq
u(x,0)=20x^2(1-x) e^{-5x}, \;\; p(x,0)=(x-0.5)e^{2x}
\eeq
Combining polynomials and exponents we ensure that different trigonometric modes are present in the spectrum of initial data that leads to a rich spectrum in time. 

First of all, the control of just two coefficients in expressions for derivatives   is no longer able to ensure non growing cost function beyond the  assimilation window. We see in \rfg{exp-costfn}A that the cost function of the model with optimal coefficients  $\alpha_0\; , \alpha_1$ grows after the assimilation end in the same way as  the cost function of the original model. Solid and upper dashed (that corresponds to $J=1$) lines  are parallel to each other. In fact, the data assimilation reduces the model's error approximately 20 times, but the behavior of the error remains the same. Consequently, we can not state that the model's error with optimal boundary approximation will always be small. Increasing with time, the error will later reach  the same values as the error of the original model.

\figureleft{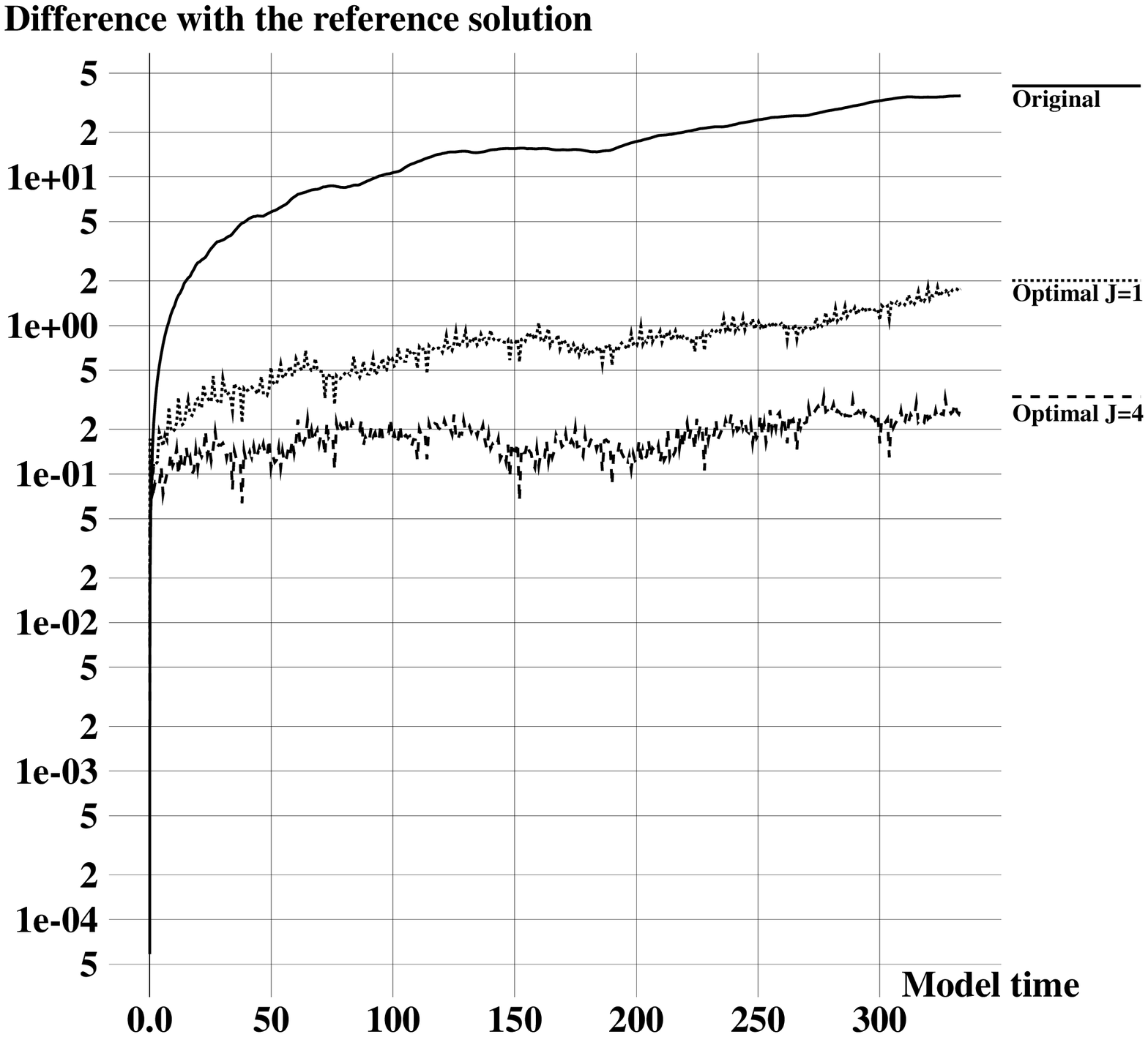}{ Cost function of numerical solutions  with $J=1$ and $J=4$. Solutions with the original boundary scheme  are plotted with solid lines, with optimal scheme -- with dashed lines.  } {exp-costfn}
\figureright{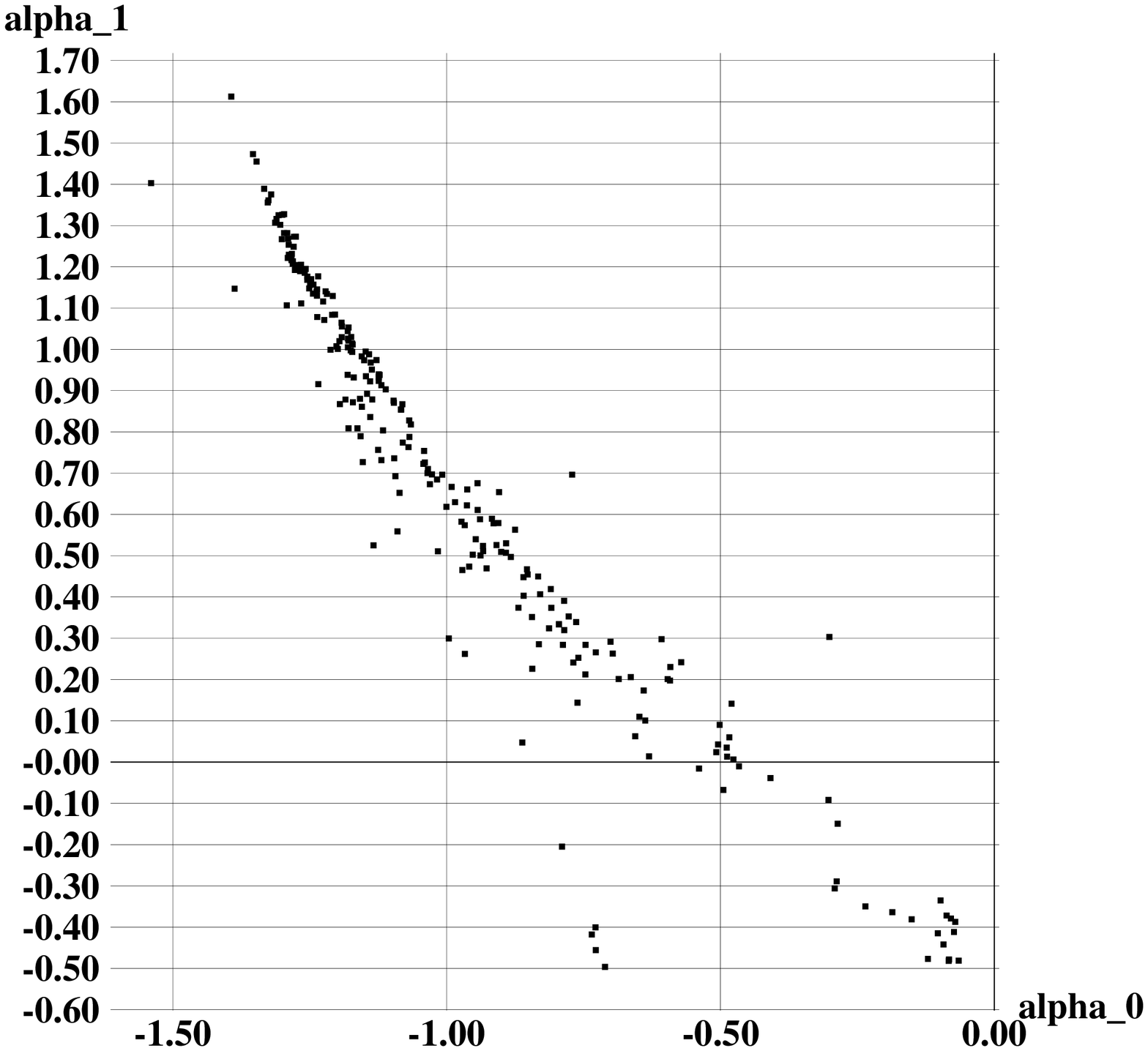}{ First two optimal coefficients for $\der{u}{x}$ with $J=4$ in experiments with different assimilation windows.   }

This fact can  be explained by the analysis of the expression  \rf{hmod} for the second order scheme
\beq
h_{modified}= h-\fr{Nh}{2}\fr{\beta-1}{\beta} = h - \fr{\beta-1}{2\beta} = h+\biggl(\fr{1}{2}-\fr{\tau\; \sin(kh/2)}{h\; \sin(k\tau)}\biggr).
\eeq
The coefficient in the expression for   the derivative of $u$ becomes
\beq
 c=\fr{h}{h_{modified}}= \fr{h^2 \sin(k\tau)}{(h^2-h/2)\sin(k\tau)+\tau \;\sin(kh/2)}
\eeq
For the given parameters ($h=1/30,\; \tau=1/120$) we get 
$
c=\fr{1}{15 \cos(k/120)-14}
.$
The denominator of this expression   vanishes and changes sign  when $k=120 \arccos(\fr{14}{15}) \sim 14.026\pi$. Consequently, optimal expression for $\der{u}{x}$ at the first point for the wave with $k=15\pi$ must have an opposite sign with respect to the classical approximation, namely: $\der{u}{x}\biggl|_{1/2}=-7.05 \fr{u_1-u_0}{h}$. The wave with $k=15\pi$ is present in the spectrum of initial conditions (its  wavelength is equal to $\fr{2\pi}{k}=4h$) but corresponding  optimal expression for the derivative can not be obtained in the assimilation procedure because the scheme is instable with negative $c$. Hence, the minimum is unreachable and we can not obtain the optimal approximations of derivatives near the boundary. Data assimilation allows us to compensate the error in wave velocities for first 14 trigonometric modes, but all other modes continue to  propagate with wrong velocities. That's why the cost function in the  experiment with assimilated data is smaller than the original cost function, but the long time behavior is similar in both experiments. 

In order to obtain the cost function that does not increase after the end of assimilation, we may try to control more coefficients $\alpha$ in \rf{bndsch} in order to be able to identify  optimal coefficients  in the domain where the scheme is stable. Increasing the number of controlled parameters, we increase the number of degrees of freedom and the dimension of the kernel of Hessian. The intersection of the kernel and the region where the scheme is stable may become non null and allow the assimilation  to reach the minimum. 

Indeed, if we perform assimilation with $J=4$, i.e. 5 coefficients $\alpha$ in \rf{bndsch}, we get smaller non increasing cost function (lower dashed line in \rfg{exp-costfn}A).  
 
Coefficients $\alpha^u$ in the experiment with $J=4$ are  distributed in  a wide area, showing larger multidimensional  kernel of the Hessian. An example of such a distribution is shown in \rfg{exp-costfn}B. To obtain this figure, we perform a set of experiments with different assimilation windows in range from 800 to 5000 time steps of the model. In each assimilation we get different sets of coefficients $\alpha$ but almost the same final cost function showing all obtained $\alpha$ are in the kernel of the Hessian.   Only the first two coefficients are plotted in \rfg{exp-costfn}B. One can see, they occupy much wider area than in experiments with one or two trigonometric waves and $J=1$ shown in \rfg{2wv}B.

 \section{Conclusion}

The purpose of this paper is to study the 
variational data assimilation procedure applied for identification of the optimal parametrization of the derivatives near the boundary on the example of   a simple wave equation in view to use this kind of data assimilation in ocean models.  Consequently,  conclusions are formulated from this point of view. 

 Comparing this procedure with now well developed data assimilation intended to identify optimal initial data, we can say there are both common points and differences as well.  
 
 Tangent \rf{tlm} and  adjoint  \rf{matadj} models    are composed by two  terms, presented by  \rf{D} and \rf{hats}.  The first one, $D^{(u)}$ \rf{D}, governs   the evolution of a small perturbation by the model's dynamics. This term is common for any data assimilation no matter what parameter we want to identify.  The second one, $\hat U$ or $\hat P$, \rf{hats},  determines the way how the uncertainty is introduced into the model. So, if we intend to identify an optimal boundary parametrization for a model with an existing adjoint developed for data assimilation and identification of initial point, we can use this adjoint as \rf{D} part because this part is common for any data assimilation.  However, the part decribed by \rf{hats} must be developed from the beginning because it is specific to the particular control parameter. 
This development may be technically difficult for complex models, especially on grids with distributed variables like Arakawa's "C"-grid.  Numerous interpolation and differentiation operators are frequently applied successively to a model's variable on these grids resulting in nonlinear dependence of the model's state on control coefficients. Development of the adjoint model and, particularly, it's \rf{hats} part, is complicated by working with nonlinearities of higher degree.        

Another difference consists in the number of control parameters and their dimensions. The dimension of initial point of the model is usually equal to the dimension of the model's state variable. Contrary to this, when we  control boundary parametrization, the dimension of control variables is very different from the dimension of the model's variable.  Moreover, the dimension of the control might be lower than the dimension of the model state because  the dimension of the control  is proportional to the length of the boundary of the domain, while the dimension of the model's state relates to the area of the domain. That means the quantity of controlled parameters and the dimension of the gradient of cost function may be much lower than the quantity of variables in the models state. Taking into account  mentioned technical difficulties in development of the adjoint, it may be reasonable to try to calculate the gradient by some other method beginning with  the simplest finite difference method. Of course, this will be more expensive computationally, but the gain in the development procedure may compensate this excessive computational cost.

Concerning the data assimilation results, we see the data assimilation can correct errors of numerical scheme by controlling approximations near boundaries. This fact may be very useful in applications of this method to the ocean models. In addition to natural corrections of the position of the rigid boundary and prescribed physical boundary conditions, we may hope to be also able  to improve the quality of the scheme that is used in internal points.    

We can see in these assimilation experiments the presence of a kernel of the Hessian.  Consequently, the choice of optimal boundary parametrization is not unique. However, all sets of coefficients $\alpha$ from the kernel are equivalent: they provide the same (or almost the same) cost function's value and almost the same evolution of the model's solution after the end of assimilation. In the same time, we can note that optimal parametrization of derivatives near the boundary may approximate nothing in  classical sense, i.e. it may not be  valid for an arbitrary function. We have seen here that obtained expression  for $\der{p}{x}$ is valid for the cosine-type functions  with appropriate wavelength only. Hence, we must take into account that  coefficients found by data assimilation are valid for given model's parameters only. 

In the last experiment in this paper, with the wave composed by multiple trigonometric modes, we have encountered the necessity to increase the number of control parameters. In the case when the optimum is unreachable,  increasing the kernel dimension allows to obtain better results.  Combining the number of controlled coefficients (that increases the kernel dimension) and the possibility to dump the first term of the Taylor development of the resulting expression by \rf{dumpcnst} (that decreases the kernel dimension) may help us to get a reasonable result. 


\mkpicstoend 

\end{document}